\documentclass[12pt,preprint,eqsecnum]{aastex}

\newcommand*{\eg}{e.g.,\ }
\newcommand*{\ie}{i.e.,\ }
\newcommand*{\eqref}[1]{(\ref{#1})}

\slugcomment{Preprint no.\ NSF-KITP-04-127; submitted to ApJ July 22, 2004; accepted January 5, 2005}

\shorttitle{3D Vortices in PPD}
\shortauthors{Barranco \& Marcus}

\begin{document}

\title{Three-Dimensional Vortices in Stratified Protoplanetary Disks}

\author{Joseph A. Barranco\altaffilmark{1}}
\affil{Kavli Institute for Theoretical Physics, Kohn Hall, \\
University of California, Santa Barbara, CA 93106}
\email{barranco@kitp.ucsb.edu}
\altaffiltext{1}{NSF Astronomy \& Astrophysics Postdoctoral Fellow}

\and

\author{Philip S. Marcus}
\affil{Department of Mechanical Engineering, Etcheverry Hall,\\
University of California, Berkeley, CA 94720}
\email{pmarcus@kepler.berkeley.edu}

\begin{abstract}
We present the results of high-resolution, three-dimensional (3D) hydrodynamic
simulations of the dynamics and formation of coherent, long-lived vortices in
stably-stratified protoplanetary disks.  Tall, columnar vortices that extend vertically
through many scale heights in the disk are unstable to small perturbations; such
vortices cannot maintain vertical alignment over more than a couple scale heights
and are ripped apart by the Keplerian shear.  Short, finite-height vortices that
extend only one scale height above and below the midplane are also unstable,
but for a different reason: we have isolated an antisymmetric (with respect to
the midplane) eigenmode that grows with an $e$-folding time of only a few
orbital periods; the nonlinear evolution of this instability leads to the
destruction of the vortex.  Serendipitously, we observe the formation of 3D vortices
that are centered not in the midplane, but at one to three scale heights above and
below.  Breaking internal gravity waves create vorticity; anticyclonic regions of vorticity
roll-up and coalesce into new vortices, whereas cyclonic regions shear into thin
azimuthal bands.  Unlike the midplane-centered vortices that were placed \textit{ad hoc}
in the disk and turned out to be linearly unstable, the off-midplane vortices form naturally
out of perturbations in the disk, and are stable and robust for many hundreds of orbits.
\end{abstract}


\section{INTRODUCTION}

Like the atmosphere of Jupiter, protoplanetary disks are characterized
by rapid rotation and intense shear, inspiring proposals that disks may
also be populated with long-lived, robust storms analogous to the
Great Red Spot \citep{barge95,aw95}.  Such vortices may play key
roles in the formation of stars and planets:
(1) in cool, neutral protoplanetary disks, vortices might transport angular
momentum radially outward so that mass can continue to accrete onto the
growing protostar;
(2) models of protoplanet migration (that account for ``Hot Jupiters,'' gas
giant planets that closely orbit their parent stars with periods of only a few
days) are highly sensitive to the effective ``viscosity'' or turbulent transport
within disks;  transport mediated by coherent vortices rather than turbulent
eddies could lead to qualitative and quantitative changes to migration rates;
and (3) vortices rapidly sweep-up and concentrate dust particles, which may
help in the formation of kilometer-size planetesimals, the ``building blocks''
of planets, either by increasing the efficiency of binary agglomeration or
by triggering a local gravitational instability.

\subsection{The angular momentum problem in accretion disks}

One of the most vexing problems in astrophysics is how mass and angular
momentum are transported in accretion disks.  Because disks are
differentially rotating, one might think that viscous torquing could
transport angular momentum outward and mass inward onto the central object.
However, if the source of viscosity is molecular collisions, then the
time-scale for such viscous transport would exceed the age of the universe by
many orders of magnitude. \citet{shakura73} proposed that turbulence might
enhance the transport, and so introduced a turbulent or ``eddy''
viscosity: $\nu_{t} = \alpha Hc_s$, where $H$ is the pressure scale height of
the disk, $c_s$ is the sound speed, and $\alpha$ is a dimensionless number
(less than unity because turbulent eddies would most likely be subsonic and no
larger than the scale height of the disk).  The source of this turbulence
(\eg shear instabilities, convection, finite-amplitude perturbations,
magnetic fields), to say the least, has been highly controversial.
See \citet{balbus98} and \citet{stone00} for thorough reviews.

\citet{balbus91} \citep[also][]{hawley91} applied the magneto-rotational
instability (MRI) of \citet{velikhov59} and \citet{chandra60,chandra61}, and
demonstrated that weak magnetic fields can destabilize the Keplerian shear in
an accretion disk, leading to the creation of turbulence and the outward
transport of angular momentum. However, the relatively cool and nearly neutral
accretion disks around protostars most likely lack sufficient coupling between
matter and magnetic fields \citep{blaes94}, except perhaps in thin surface
layers that have been ionized by cosmic rays or protostellar x-rays
\citep{gammie96}.

Thus, there still remains strong interest in finding a purely hydrodynamic
means of mass and angular momentum transport.
\citet{bracco98}, using two-dimensional, incompressible fluid
dynamics, have shown that long-lived, coherent anticyclonic vortices emerge
out of a Keplerian shear flow that is seeded with small-scale vorticity
perturbations; they observed smaller vortices merging to form larger vortices,
demonstrating the ``inverse cascade'' of energy from small to large scales
that is a hallmark of two-dimensional turbulent flows \citep{provenzale99}.
\citet{godon99b,godon00} confirmed this result with two-dimensional,
compressible, barotropic simulations.
\citet{lovelace99} have found a linear instability of nonaxisymmetric Rossby
waves in thin, non-magnetized Keplerian disks when there is a local extremum in
an entropy-modified potential vorticity.
\citet{li00,li01} showed that such Rossby waves break and coalesce to form
vortices which radially migrate and transport mass through the disk.
\citet{klahr03} have demonstrated that a globally unstable radial entropy
gradient in a protoplanetary disk generates Rossby waves, vortices and
turbulence that may result in local outward transport of angular momentum.
To date, there are still unresolved issues regarding realistic formation
mechanisms and long-term maintenance of vortices in disks.

\subsection{The role of turbulence and vortices in planet formation}

Whether protoplanetary disks are filled with turbulent eddies or long-lived
coherent vortices (or both) is critical for not only accretion, but also
for two key processes in planet formation: protoplanet migration and
planetesimal formation.  Extrasolar planet surveys reveal a class of gas
giant planets that closely orbit their parent stars with periods of only a
few days \citep{mayor95,marcy98,marcy00}.  These ``Hot Jupiters'' most likely
did not form \textit{in situ}, but instead formed farther out in their disks
where it was cooler, and then migrated inward to their present locations
\citep{lin96b}.  The exact mechanism for this migration depends on the size of
the protoplanet: a small protoplanet ($\lesssim 10 M_{\oplus}$) raises
tides in the disk which exert torques back on the protoplanet, causing it to
migrate inward (Type I migration), whereas a larger protoplanet opens up a
gap in the disk, and both gap and protoplanet migrate inward together on a slow
``viscous'' timescale (Type II migration) \citep{ward97}.  In numerical
simulations of protoplanet migration, the viscous evolution is usually modeled
with an eddy viscosity.  Although such a crude turbulence model is useful for
estimating global disk properties, it is not at all clear that it is
appropriate to apply it to complex disk-planet interactions.  Does a turbulent
disk really behave exactly the same way as a laminar disk with just a larger
viscosity?  Qualitatively and quantitatively, transport and mixing due to
long-lived coherent vortices are not necessarily well modeled by an enhanced
viscosity \citep{koller03}.

Another aspect of planet formation that is extremely sensitive to turbulence
is the formation of planetesimals, the kilometer-size ``building blocks'' of
planets.  In a quiescent, laminar protoplanetary disk, planetesimals may
form directly from the gravitational clumping of a thin, dense dust sublayer
that has settled into the midplane \citep{safranov60,goldreich73}.  However,
turbulence (either from MRI, or generated by Kelvin-Helmholtz-like
instabilities of the vertical shear between the dust sub-layer in the midplane,
which orbits at the Keplerian velocity, and the gas-rich layers immediately
above and below it, which orbit at a slightly slower rate because of the
outward gas pressure force) might ``kick up'' the dust and prevent the
sub-layer from reaching the critical density necessary for gravitational
instability \citep{cuzzi93,weidenschilling95,champney95}.  Further growth of
grains must instead continue via binary agglomeration
\citep{weidenschilling93}.

Vortices may aid in the formation of planetesimals in either scenario.
\citet{barge95}, \citet{tanga96}, \citet{chavanis00}, and
\citet{fuentemarcos01} tracked the motion of individual particles in
analytic models of two-dimensional anticyclones embedded in Keplerian shear.
\citet{klahr97} considered simple models of vortices that corresponded to
convective cells.  
\citet{johansen04} used two-fluid simulations to study the interactions of the
gas and dust components inside vortices.  However, because of the relatively
low resolution and high numerical dissipation in their simulations, their
vortices were short-lived and decayed away in roughly one orbit.
In all these models, gas drag caused grains to spiral
into the vortex centers.
Density enhancements inside vortices might increase the efficiency of
binary agglomeration, or might trigger \textit{local} gravitational clumping
within the vortex center as opposed to the global gravitational clumping of
the whole dust sub-layer envisioned in the original \citet{safranov60} and
\citet{goldreich73} theory.

\subsection{Outline}

This work is the first in a series on high-resolution numerical simulations
of the dynamics and formation of long-lived, coherent vortices in
protoplanetary disks, and on what role they may play in star and planet
formation.
In \S 2, we will overview dimensional analysis and key timescales
for vortices in protoplanetary disks.  We will discuss the deficiencies of
2D analyses and explain the need for 3D simulations.
In \S 3, we will present the hydrodynamical equations for the base disk and
the vortex flow, as well as briefly describe the numerical method. 
Results of numerical simulations will be presented in \S 4, including
a discussion on angular momentum transport.
Conclusions and avenues for future work appear in \S 5.  

\section{A PRIMER FOR VORTEX DYNAMICS IN DISKS}

\subsection{Dimensional analysis and key timescales}

Protoplanetary disks are cool in the sense that the gas sound speed $c_s$ is
much slower than the Keplerian orbital velocity $V_K(r) \equiv r\Omega_K(r)
\equiv \sqrt{GM_{\star}/r}$, where $G$ is the gravitational constant,
$M_{\star}$ is the protostellar mass, and $r$ is the cylindrical radius to the
protostar.   Hydrostatic balance implies that the time it takes sound waves to
traverse the thickness of the disk is of order the orbital period.  Thus, cool
disks are geometrically thin \citep[see][]{frank85}:
\begin{mathletters}
\begin{eqnarray} \label{E:hydrostatic}
c_s & \sim & \Omega_K H, \label{E:hydrostatic1}\\
\delta \equiv c_s/V_K & \sim & H/r < 1,\label{E:coolthin}
\end{eqnarray}
\end{mathletters}
where $H$ is the vertical pressure scale height.
The radial component of the protostellar gravity nearly balances the
centrifugal force, but because of the relatively weak outward radial
pressure force, the gas orbits at slightly slower than the Keplerian
velocity:
\begin{equation}\label{E:Keplerian}
\Omega(r,z) = \Omega_K(r)\left[1-\eta(r,z)\right],
\end{equation}
where $\eta\sim\mathcal{O}(\delta^2)$.
The horizontal shear rate of this base flow is:
\begin{equation}\label{E:shear}
\sigma_K \equiv r\frac{\partial\Omega_K}{\partial r} =
-\frac{3}{2}\Omega_K.
\end{equation}
Keplerian shear is anticyclonic (as indicated by the negative sign), and is
comparable in magnitude to the rotation rate itself.  As we will see, this has
important consequences for the properties of vortices in protoplanetary disks.

Shear will stretch and tear a vortex apart unless: (1) the vortex rotates
in the same sense as the ambient shear, and (2) the vortex is at least as
strong as the shear \citep{ms71,kida81,marcus90a,marcus93}.
The first condition implies that all long-lived vortices in a protoplanetary
disk will be anticyclones; that is, the vortices will rotate in a sense
opposite of that of the overall rotation of the disk.
Let $\tilde{v}_{\bot}$ be the characteristic horizontal speed of gas around
a vortex with aspect ratio $\chi\equiv\Lambda_{\phi}/\Lambda_r>1$, where
$\Lambda_r$ (along the radial direction in the disk) is the minor axis and
$\Lambda_{\phi}$ (along the azimuthal direction in the disk) is the major axis.
The characteristic $z$-component of vorticity\footnote{Recall that vorticity is
defined $\mathbf{\omega}\equiv\mathbf{\nabla}\times\mathbf{v}$, and is
\textit{two times} the local angular velocity of the fluid.}
in the core of the vortex is
$\tilde{\omega}_z\sim\tilde{v}_{\bot}/\Lambda_r$.
The second condition above implies that the horizontal speed of gas around
the vortex must be of order the differential velocity across the vortex due to
the ambient shear, or equivalently, that the characteristic vorticity is of
order the shear rate:
$\tilde{\omega}_z\sim\tilde{v}_{\bot}/\Lambda_r\sim\sigma_K.$
\citet{ms71} and \citet{kida81} analytically determined the exact
steady-state solution for a 2D elliptical vortex of uniform vorticity
embedded in a uniform shear and showed that:
\begin{equation} \label{E:moore_saffman}
\frac{\tilde{\omega}_z}{\sigma} = \left(\frac{\chi+1}{\chi-1}\right)\frac{1}{\chi},
\end{equation}
for a vortex that rotates in the same sense as the shear.
A vortex that is weak relative to the shear will be stretched out and have
a large aspect ratio, whereas a vortex that is comparable in strength
to the shear will be more compact and have an aspect ratio closer to unity.
Because the Keplerian shear is comparable in magnitude to
the Keplerian rotation rate, the rotational period of 
gas around the core of the vortex, $\tau_{vor}\equiv 4\pi/\tilde{\omega}_z$,
is of order the orbital period of the vortex around the protostar,
$\tau_{orb}\equiv 2\pi/\Omega$.
The Rossby number is defined to be the ratio of these two timescales:
\begin{equation}\label{E:Rossby}
Ro \equiv \frac{\tilde{\omega}_z}{2\Omega}
\sim \frac{\tau_{orb}}{\tau_{vor}}
\sim \frac{3}{4}\frac{\chi+1}{\chi-1}\frac{1}{\chi},
\end{equation} 
which is order unity for compact vortices.
In contrast, the Great Red Spot (GRS) on Jupiter rotates with a period of
roughly six days, whereas the planet rotates in just under 10 hours, implying
a Rossby number $Ro \approx 0.18$ \citep{marcus93}.\footnote{For a rotating
planet, $Ro\sim(\tau_{orb}/\sin\lambda)/\tau_{vor}$, where $\lambda$ is
the latitude.  Only the component of the rotation normal to the surface
of the planet enters into the horizontal component of the Coriolis force.}  

As we are interested in long-lived vortices, we require that the gas velocity
be subsonic; otherwise, sound waves and shocks would rapidly dissipate the
kinetic energy of the vortex.   Equations \eqref{E:hydrostatic} and
\eqref{E:Rossby} imply:
\begin{equation} \label{E:mach1}
\epsilon \equiv \frac{\tilde{v}_{\bot}}{c_s} \sim \frac{\Lambda_r}{H}Ro,
\end{equation}
where $\epsilon$ is defined to be the horizontal Mach number.
\textit{Thus, the horizontal extent of subsonic, compact vortices in a Keplerian
shear cannot be much greater than the scale height of the disk.}
Such vortices are 3D in nature. 
In contrast, the GRS has considerably more ``pancake-like'' dimensions:
26,000~km $\times$ 13,000~km $\times$ 40~km \citep{marcus93}.

Another important timescale is the Brunt-V\"{a}is\"{a}l\"{a} period:
$\tau_{BV}\equiv 2\pi/\omega_{BV}$, which is the period of oscillations
in a convectively stable atmosphere (\eg internal gravity waves).
The Brunt-V\"{a}is\"{a}l\"{a} frequency is \citep[see][]{pedlosky79}:
$\omega_{BV}\equiv\sqrt{(g/C_p)\partial{\bar{s}}/\partial z}$,
where $g$ is the local gravitational acceleration, $C_p$ is the specific heat
at constant pressure, and $\bar{s}$ is the mean entropy. If the protoplanetary
disk is vertically isothermal, then $\omega_{BV}\approx\Omega_K|z|/H$.
That is, except in the immediate vicinity of the midplane, the
Brunt-V\"{a}is\"{a}l\"{a} period is also of order the orbital period in the
disk.  The internal Froude number is defined as the ratio of the
Brunt-V\"{a}is\"{a}l\"{a} period to the rotational period of the gas in a
vortex:
\begin{equation}
Fr \equiv \frac{\tilde{\omega}_z}{2\omega_{BV}}
\sim \frac{\tau_{BV}}{\tau_{vor}} \sim \left(\frac{H}{z}\right)Ro.
\end{equation}
When the Froude number is much less
than unity, a system is considered to be strongly stratified;
large-scale vertical motions are suppressed and the different layers of the
fluid are only weakly coupled.

Table \ref{T:timescales} summarizes the values of the key timescales discussed
so far.  For comparison, values for the GRS are presented as well.  On
Jupiter, these timescales are well ordered:
$\tau_{vor}\gg\tau_{orb}\gg\tau_{BV}$; whereas in a protoplanetary disk:
$\tau_{vor} \sim \tau_{orb} \sim \tau_{BV}$.  The near-equality of timescales
in a protoplanetary disk simply follows from the fact that at any given
location in the disk, the protostellar gravity sets the orbital velocity,
shear, and strength of the stratification.  Because the
Brunt-V\"{a}is\"{a}l\"{a} frequency for the GRS is so much faster than the
other timescales, the vertical and horizontal dynamics can be decoupled; the
GRS is well-modeled with two-dimensional fluid dynamics.   No such clear
separation of timescales occurs for a protoplanetary disk, and the horizontal
and vertical dynamics cannot be decoupled.  In general, when the Rossby and
Froude numbers are small (\eg rotation and stratification dominate
over shear and nonlinear advection), then the horizontal and vertical motions
can usually be decoupled and two-dimensional analyses can safely be applied. 

In many of the common 2D models (\eg ``shallow-water'', quasigeostrophic, or
2D-barotropic),  vorticity (or a potential vorticity) is an advectively
conserved quantity (because baroclinicity and vortex tilting are
neglected).  In 2D, only the $z$-component of vorticity
is nonzero, and vortex lines can only be oriented vertically.  The rotation
axes of long-lived coherent vortices as well as transient turbulent eddies
are constrained to be aligned parallel (or anti-parallel) to the vertical
axis.  Furthermore, shear will stretch out and destroy those vortices that
do not have the same sense of rotation as the shear.  Thus, a 2D shear flow
will be populated with vortices that are all perfectly aligned vertically and
with the same sense of rotation.  It has been demonstrated in both laboratory
experiments  and numerical simulations that in such restricted flows, 
small regions of vorticity merge to form larger vortices --- this phenomenon
is called an ``inverse cascade'' of energy from small to large scales
\citep{kraichnan67,lesieur97,paret98,baroud03}
Out of an apparently chaotic flow filled with small-scale, transient eddies,
large-scale coherent features can emerge.

In 3D, the dynamics of turbulence are very different.  
Vortex lines can be oriented in any direction, and they can twist and bend.
Vortices and eddies tilt and stretch their neighbors, disrupting them and
eventually destroying them as they give up their energy to smaller eddies.
In fact, this is the foundation for the Richardson and Kolmogorov model of
3D, isotropic turbulence: turbulent kinetic energy is transfered (via
nonlinear interactions) from large to small to smaller eddies on down
(\ie ``forward cascade'') until it is destroyed by viscous dissipation.

An open question (and a very active area of basic research in fluid dynamics)
is what happens when rotation and stratification are important, but not
overwhelmingly dominant (\eg when the Rossby and/or Froude numbers are of order
unity)?  How much rotation and stratification are necessary for a real, 3D
flow to exhibit 2D characteristics, such as inverse cascades? 
\citet{bracco98} and \citet{godon99b,godon00} showed that small vortices
merged to form larger vortices in 2D simulations, but these results have
yet to be confirmed with 3D simulations.

\begin{table}
\begin{center}
\begin{tabular}{ccc} \tableline
Timescale                           & GRS             & PPD \\
\tableline
$\tau_{vor}\equiv 4\pi/\tilde{\omega}_z$ & 8 d.    & $\sim$ 1 y.\\
$\tau_{orb}\equiv 2\pi/\Omega$                & 10 h. & $\sim$ 1 y.\\
$\tau_{BV} \equiv 2\pi/\omega_{BV}$      & 6 m.  & $\sim$ 1 y.\\
\tableline
$Ro \equiv \tau_{orb}/\tau_{vor}$  & 0.18                        & $\mathcal{O}(1)$ \\
$Fr \equiv \tau_{BV}/\tau_{vor}$    & $5\times 10^{-4}$ & $\mathcal{O}(1)$ \\
$Ri \equiv 1/Fr^2$                            & $4\times 10^{6}$  & $\mathcal{O}(1)$ \\
\tableline
\end{tabular}
\caption{\label{T:timescales}
Comparison of important dynamical timescales for the Great Red Spot
and a vortex in a protoplanetary disk.  Here, we use ``year'' in the more
general sense to refer to the orbital period in the disk.
$\tau_{vor}$ is the orbital period of gas around the core of the vortex,
$\tau_{orb}$ is the rotational period of the system, and
$\tau_{BV}$ is the Brunt-V\"{a}is\"{a}l\"{a} period, or the timescale for
vertical thermal oscillations in a stably stratified atmosphere.
Note that for the GRS,
$\tau_{vor}\gg\tau_{orb}\gg\tau_{BV}$, whereas for the PPD,
$\tau_{vor} \approx \tau_{orb} \approx \tau_{BV}$.  The lower half of the table
expresses the same data in terms of common nondimensional parameters from
fluid dynamics: the Rossby number, the internal Froude number, and
the Richardson number.
For a planetary atmosphere, only the component of the rotation vector normal to
the surface is relevant, so the Rossby number is larger by a factor
$1/\sin\lambda$, where $\lambda$ is the latitude on the planet.}
\end{center}
\end{table}

\subsection{A model of a 3D vortex}

\begin{figure}
\epsscale{0.6}
\plotone{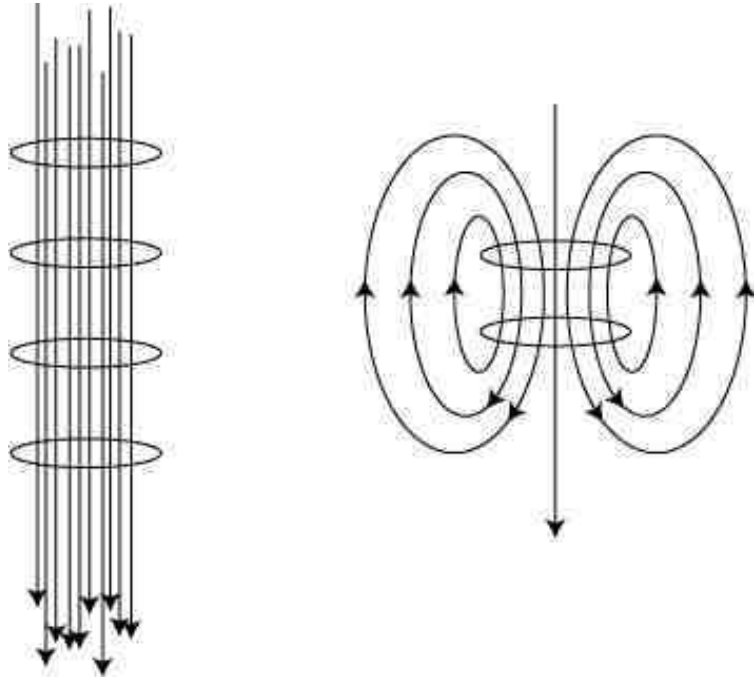}
\caption{\label{F:vortex_lines} Vortex lines in an infinite columnar
vortex and a finite-height cylindrical vortex.  Downward oriented arrows
denote anticyclonic vorticity, whereas upward-oriented arrows indicate
cyclonic vorticity.}
\end{figure}

\begin{figure}
\epsscale{1.0}
\plotone{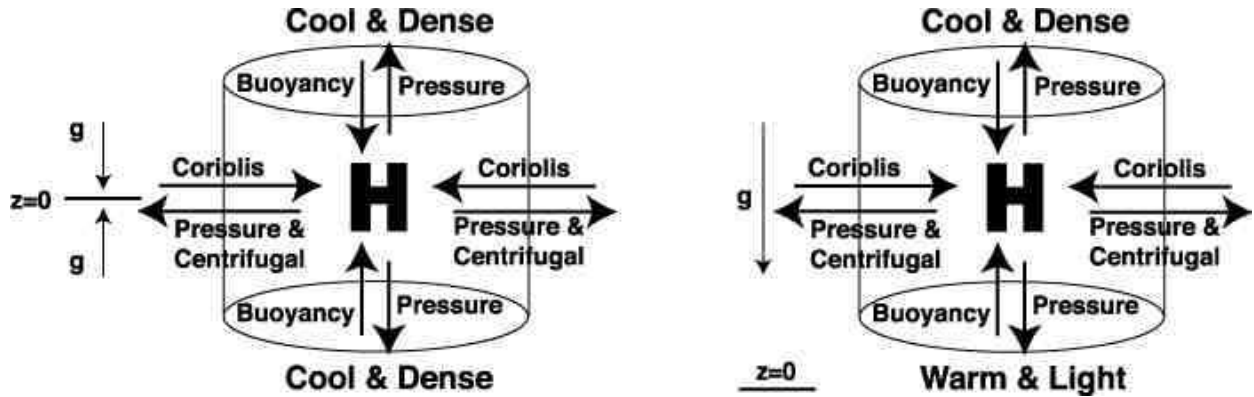}
\caption{\label{F:force_balance} Balance of forces in an anticyclone with
Rossby number less than unity.}
\end{figure}

\begin{table}
\begin{center}
\begin{tabular}{cc} \tableline
Variable                                                   & Scaling        \\
\tableline
$\tilde{v}_{\bot}/c_s$                                   & $\epsilon\sim(\Lambda_r/H)Ro$     \\
$\tilde{p}/\bar{p} \sim \tilde{\rho}/\bar{\rho} \sim \tilde{T}/\bar{T}$& $\epsilon^2/Ro$ \\
$\tilde{v}_z/c_s$                                   & $<\epsilon^2$   \\
\tableline
\end{tabular}
\caption{\label{T:scalings} How velocity and thermodynamic variables
scale with Mach number $\epsilon$ and Rossby number $Ro$.}
\end{center}
\end{table}

One of the best ways to visualize a vortex in three dimensions is to graph
vortex lines, curves that are everywhere tangent to the vorticity vector
field.  Vorticity is, by definition, divergence-free; thus, vortex lines (like
magnetic field lines) cannot have beginnings or endings within the fluid,
but must extend to the boundaries, or off to infinity, or form closed loops.
The simplest example of a 3D vortex is an infinite column of rotating
fluid in which the fluid velocity is independent of height.  The core of
uniform vorticity is threaded by infinitely-long parallel vortex lines
(see Figure \ref{F:vortex_lines}a).
One might imagine chopping off the ends of an infinite column to create a
finite-height cylinder of rotating fluid.  However, vortex lines cannot
have loose ends in the fluid, and instead must wrap-around and form closed
loops (see Figure \ref{F:vortex_lines}b).  Note that in the core of such a
vortex, the vortex lines will be oriented in one direction, whereas in a halo
surrounding the core, the vortex lines will be oriented in the opposite
direction.  

We now turn to the balance of forces in a 3D vortex.  In any horizontal
plane, the centrifugal force always points radially outward from the vortex
center, whereas the Coriolis force points outward for cyclones and inward
for anticyclones.   When the Rossby number is much greater than unity, the
Coriolis force is negligible.  The outward centrifugal force must be balanced
by an inward pressure force; such vortices must have low-pressure
cores.   When the Rossby number is less than unity, the centrifugal
force is small and the Coriolis force is balanced by the pressure force. 
Cyclones must have low-pressure cores and anticyclones must have
high-pressure cores.

In the vertical direction,
the only force that can balance the pressure force is buoyancy.  
Figure~\ref{F:force_balance}a shows the balance of
forces in a low Rossby number anticyclone that is vertically centered on
the midplane, whereas figure~\ref{F:force_balance}b shows the same for
an anticyclone located completely above the midplane.
In the first case, the high-pressure anticyclone must have cool, dense lids
to provide a buoyant force toward the midplane that balances the pressure
force away from the midplane.  In the second case, the top lid (\ie the one
farthest from the midplane) must be cool and dense, whereas the bottom lid
(\ie the one closest to the midplane) must be warm and light.
These thermal lids will be weakened or destroyed either by radiation of heat
or turbulent erosion.  Thus, a vertical flow must develop within the
vortex to maintain the lids.  For a high-pressure anticyclone centered on the
midplane, gas will rise away from the midplane, adiabatically expanding
and cooling.  At the lids, the gas diverges outward and recirculates back
toward the midplane along the edges of the vortex.  For an anticyclone that
is completely above the midplane, gas will rise (sink) toward the top (bottom)
lid, where it will expand and cool (compress and heat-up).

We can use the arguments in this section to derive simple scaling relationships
for the vortex flow variables, which will then guide us in making further
approximations to the fluid equations.  We decompose each flow
variable into a time-independent, axisymmetric component describing the base
disk flow, which we will denote with overbars, and a time-dependent component
describing the vortex flow, which we will denote with tildes
(\eg $\rho = \bar{\rho}+\tilde{\rho}$, $p = \bar{p}+\tilde{p}$).
As discussed in the previous subsection, we assume that the horizontal
vortex flow is subsonic: $\tilde{v}_{\bot}/c_s\sim\epsilon<1$.   If the Rossby
number is of order unity or less, then the horizontal pressure force must be of
the same order as the Coriolis force (geostrophic balance):
$\tilde{p}/\bar{\rho}\Lambda_r\sim 2\Omega\tilde{v}_{\bot}$, where
$\tilde{p}$ is the pressure perturbation associated with the vortex,
and $\bar{\rho}$ is the mean density.  Using \eqref{E:hydrostatic} and \eqref{E:mach1},
one can show that the pressure perturbation in the vortex must scale as:
$\tilde{p}/\bar{p} \sim \epsilon^2/Ro$.
We also assume the vertical buoyancy force is of order the vertical pressure
gradient: $g_z\tilde{T}/\bar{T}\sim\tilde{p}/\bar{\rho}H$, which leads to the
temperature fluctuation having the same scaling as the pressure
fluctuation: $\tilde{T}/\bar{T}\sim\tilde{p}/\bar{p}$.  Because of the ideal gas law,
this implies that density fluctuations also have the same scaling.
The scaling for the vertical velocity is set by the maintenance of temperature
perturbations in the thermal lids.  In the temperature equation, temperature fluctuations
are created via pressure-volume work: $\tilde{v}_z\bar{T}(d\bar{s}/dz)/C_v$, where
$\tilde{v}_z$ is the vertical velocity, $\bar{s}$ is the mean entropy profile, and $C_v$
is the heat capacity at constant volume.
The destruction process is either radiative heating/cooling, $\tilde{T}/\tau_{cool}$, where
$\tau_{cool}$ is a characteristic cooling time; or turbulent advection,
$\tilde{T}/\tau_{eddy}\sim\tilde{v}_{\bot}\tilde{T}/\Lambda_r$, where $\tau_{eddy}$ is
a characteristic eddy diffusion time.
If the vortex flow is smooth and laminar, then we balance pressure-volume work
with thermal cooling to obtain the scaling: $\tilde{v}_z\sim (\epsilon^2/Ro)H/\tau_{cool}$.
Thus, if the vortex is laminar and the cooling time is long, the vertical velocity
can be quite small.  On the other hand, if the vortex flow is highly turbulent,
we balance pressure-volume work and the turbulent advection of heat to
obtain the scaling: $\tilde{v}_z\sim\epsilon^2$.
These scaling relationships are summarized in Table \ref{T:scalings}.

\section{FLUID EQUATIONS AND NUMERICAL METHOD}

\subsection{Cartesian and anelastic approximations}

Motivated by the scalings discussed in the previous section,
we focus on subsonic, compact vortices that have horizontal extent $\Lambda_r$
comparable to the vertical scale height $H$, which is much smaller than the
distance $r$ to the protostar: $\Lambda_r/H\sim\epsilon/Ro\lesssim1$ and
$\Lambda_r/r\sim\epsilon\delta/Ro\ll 1$.  This allows us to make two key
approximations to the hydrodynamics.

The first simplification we make is the Cartesian approximation
\citep{goldreich65b}: we simulate the hydrodynamics only within a small patch
of the disk $(\Delta r\ll r_0,\Delta\phi\ll 2\pi)$ that co-rotates with the gas at some
fiducial radius $r_0$.  We map this patch of the disk onto a Cartesian grid:
$r-r_0 \rightarrow x$, $r_0(\phi - \phi_0) \rightarrow y$, $z\rightarrow z$,
$v_r \rightarrow v_x$,  $v_{\phi} \rightarrow v_y$, and $v_z\rightarrow v_z$.
The background shear and mean thermodynamic variables have radial gradients
that vary on the length scale $r$ which is much larger than the characteristic size of a
subsonic vortex.  For example, let $\bar{q}$ represent any background disk variable;
then the variation $\delta\bar{q}$ over the size of a vortex is:
$\delta\bar{q}/\bar{q} \sim (\partial\ln\bar{q}/\partial r)\Lambda_r\sim \Lambda_r/r\ll 1$.
This allows us to neglect radial gradients of the mean thermodynamic variables
and to linearize the Keplerian shear flow.  The time-independent, axisymmetric base
Keplerian flow in the rotating frame, which we denote with overbars, is then:
\begin{mathletters}
\begin{eqnarray}\label{E:base_flow}
\mathbf{\bar{v}} &=& -\case{3}{2}\Omega_0 x\mathbf{\hat{y}},\label{E:base_velocity}\\
\bar{T}          &=& T_0,                     \label{E:base_temperature}     \\
\bar{\rho}       &=& \rho_0\exp(-z^2/2H_0^2), \label{E:base_density}         \\
\bar{p}          &=& p_0\exp(-z^2/2H_0^2),    \label{E:base_pressure}        \\
\bar{s}          &=& \mathcal{R}z^2/2H_0^2,   \label{E:base_entropy}  
\end{eqnarray}
\end{mathletters}
where $\bar{T}$, $\bar{\rho}$, $\bar{p}$, $\bar{s}$ are the mean
temperature, density, pressure, and entropy, respectively,
which depend only on the vertical coordinate $z$.
We have defined:
$\Omega_0\equiv\Omega_K(r_0)$, $p_0\equiv\rho_0\mathcal{R}T_0$,
$H_0^2\equiv\mathcal{R}T_0/\Omega_0^2$, and $\mathcal{R}$ is the gas constant.
The errors in making this Cartesian approximation are of order
$\mathcal{O}(\epsilon\delta,\delta^2)$. 
Note that neglecting radial gradients of the background disk properties effectively
filters out Rossby waves (which require a gradient in the background vorticity) and
large-scale baroclinic instabilities.  Our rationale for this is
that if there were indeed large-scale baroclinic instabilities, they would produce
large, supersonic vortices which would rapidly decay from radiation of 
acoustic waves and shocks. 

The second simplification we make is the anelastic approximation:
all variables are expanded in powers of the Mach number 
(\eg $\mathbf{v} = \epsilon(\mathbf{\bar{v}} + \mathbf{\tilde{v}}) + ...\;$;
$p=\bar{p} + \epsilon^2\tilde{p} + ...\;$; 
$\rho=\bar{\rho} + \epsilon^2\tilde{\rho} + ...\;$).
These expansions are then substituted into the Euler equations, and terms
are grouped by like powers of $\epsilon$;  we keep only the first-order
corrections \citep{barranco00a}.  The anelastic approximation has been used
extensively in the study of deep, subsonic convection in planetary atmospheres
\citep{ogura62,gough69} and stars \citep{gilman81,glatzmaier81a,glatzmaier81b}.
One of the consequences of this approximation is that the total density
is replaced by the time-independent mean density in the mass continuity
equation, which has the effect of filtering high-frequency acoustic waves and
shocks, but allowing slower wave phenomena such as internal gravity waves.
The errors in making the anelastic approximation are of order
$\mathcal{O}(\epsilon^2)$.
The anelastic Euler equations in the co-rotating frame are:
\begin{mathletters}\label{E:euler}
\begin{eqnarray}
\tilde{p} &=& \bar{\rho}\mathcal{R}\tilde{T} + \tilde{\rho}\mathcal{R}\bar{T},\label{E:euler_eos}\\
0 &=& \mathbf{\nabla}\cdot\left(\bar{\rho}\mathbf{v}\right),\label{E:euler_anelastic}\\
\frac{d\mathbf{v}}{dt} &=&
-2\Omega_0\mathbf{\hat{z}}\times\mathbf{v} + 3\Omega_0^2x\mathbf{\hat{x}} - \frac{1}{\bar{\rho}}\mathbf{\nabla}\tilde{p} - \frac{\tilde{\rho}}{\bar{\rho}}\Omega_0^2z\mathbf{\hat{z}},\label{E:euler_velocity}\\
\frac{dT}{dt} &=&
-(\gamma-1)T\left(\mathbf{\nabla}\cdot\mathbf{v}\right) - (T-\bar{T})/\tau_{cool}(z),\label{E:euler_temperature}
\end{eqnarray}
\end{mathletters}
where the advective derivative is defined as
$d/dt\equiv\partial/\partial t + \mathbf{v}\cdot\mathbf{\nabla}$,
and where $\gamma\equiv C_P/C_V$ is the ratio of specific heats.
Equation \eqref{E:euler_anelastic} is the crux of the anelastic
approximation: the time-derivative has dropped out of the mass
continuity equation, which effectively filters out all acoustic phenomena.
The terms on the right hand side of the momentum equation
\eqref{E:euler_velocity} are: the Coriolis force, the tidal force (the
remainder after balancing the radial component of the protostellar
gravity and the centrifugal non-inertial force),
the pressure force, and the buoyancy force.
In equation \eqref{E:euler_temperature}, the first term on the right hand side
is pressure-volume work (which is reversible) and the second term
represents a simplified model of cooling or thermal relaxation.
In the limit where the cooling time $\tau_{cool}$ is infinitely long,
one can show that the temperature equation is equivalent to
entropy being advectively conserved: $ds/dt = 0$.


\subsection{Spectral expansions and boundary conditions}

In this and the following subsection, we briefly describe the numerical method;
a more detailed presentation can be found in \citet{barranco04a}.
We solve the Euler equations~\eqref{E:euler} with a spectral method; that is, each
variable is represented as a finite sum of basis functions multiplied by
spectral coefficients \citep{gottlieborszag77,marcus86a,canuto88,boyd89}.
The choice of basis functions for each direction is guided by the corresponding
boundary conditions.
The equations are autonomous in the azimuthal coordinate $y$, so
it is reasonable to assume periodic boundary conditions in this direction.
However, the equations explicitly depend on the radial coordinate $x$ because
of the linear background shear.  We adopt ``shearing box'' boundary conditions:
$\tilde{q}(x+L_x,y-(3/2)\Omega_0L_xt,z,t) = \tilde{q}(x,y,z,t)$,
where $\tilde{q}$ represents any of $\mathbf{\tilde{v}}$, $\tilde{p}$, $\tilde{T}$, etc. 
In practice, we rewrite the equations~\eqref{E:euler} in terms of quasi-Lagrangian or
shearing coordinates that advect with the background shear
\citep{goldreich65b,marcus77,rogallo81}:
$t' \equiv  t$, $x' \equiv  x$,  $y' \equiv  y + (3/2)\Omega_{0} x t$, and $z' \equiv  z$.
In these new coordinates, the radial boundary conditions become:
$\tilde{q}(x'+L_x,y',z',t') = \tilde{q}(x',y',z',t')$.  That is, shearing
box boundary conditions are equivalent to periodic boundary conditions in the
shearing coordinates.  Physically, this means that the periodic images at different radii
are not fixed, but advect with the background shear.

In the shearing coordinates, the equations are autonomous in both $x'$ and
$y'$ (although they now explicitly depend on $t'$), making a Fourier basis the
natural choice for the spectral expansions in the horizontal directions:   
\begin{equation}\label{E:spectral}
\tilde{q}(x',y',z',t') = \sum_{\mathbf{k}} \hat{q}_{\mathbf{k}}(t')e^{ik'_xx'}e^{ik'_yy'}\phi_n(z'),
\end{equation}
where $\tilde{q}$ is any variable of interest, $\{\hat{q}_{\mathbf{k}}(t')\}$ is the
set of spectral coefficients, and $\mathbf{k}=\{k'_x,k'_y,n\}$ is the set
of wavenumbers.
We have implemented the simulations with two different sets of basis
functions for the spectral expansions in the vertical direction, corresponding
to two different sets of boundary conditions:

\noindent (i) For the truncated domain $-L_z\le z \le L_z$, we use Chebyshev polynomials:
$\phi_n(z) = T_n(z/L_z) \equiv \cos(n\xi)$, where $\xi\equiv\cos^{-1}(z/L_z)$.
We apply the condition that the vertical velocity vanish at the
top and bottom boundaries: $\tilde{v}_z(x,y,z\!=\!\pm L_z,t) = 0$.

\noindent (ii) For the infinite domain  $-\infty < z < \infty$, we use rational Chebyshev
functions: $\phi_n(z) = \cos(n\xi)$ (for $\tilde{v}_x$, $\tilde{v}_y$,
$\tilde{\omega}_z$, and all the thermodynamic variables) or $\phi_n(z)=\sin(n\xi)$
(for $\tilde{v}_z$, $\tilde{\omega}_x$ and $\tilde{\omega}_y$), where
$\xi \equiv \cot^{-1}(z/L_z)$.  In this context, $L_z$ is no longer the
physical size of the box, but is a mapping parameter; exactly one half
of the grid points are within $|z|\le L_z$, whereas the other half are widely
spaced in the region $L_z < |z| < \infty$.    No explicit boundary conditions
on the vertical velocity are necessary when we solve the equations on the
infinite domain because the $\phi_n(z)=\sin(n\xi)$ basis functions
individually decay to zero at large $z$.

\subsection{Brief description of the numerical method}

The equations are integrated forward in time with a fractional step method:
the nonlinear advection terms are integrated with an explicit second-order
Adams-Bashforth method, and the pressure step is computed with a
semi-implicit second-order Crank-Nicholson method.  The time integration
scheme is overall globally second-order accurate.  Unlike finite-difference
methods, spectral methods have no inherent grid dissipation; energy
cascades to smaller and smaller size scales via the nonlinear interactions,
where it can ``pile-up'' and potentially degrade the convergence of the
spectral expansions.  We employ a $\nabla^8$ hyperviscosity or low-pass filter
every timestep to damp the energy at the highest wavenumbers.

Different horizontal Fourier modes interact only through the nonlinear
advective terms; once these terms are computed, the horizontal Fourier
modes can be decoupled.  This motivated us to parallelize the code:
each processor computes on a different block of data in horizontal
Fourier wavenumber space.  Parallelization is implemented with
Message Passing Interface (MPI) on the IBM Blue Horizon and IBM Datastar at
the San Diego Supercomputer Center, typically using between 64 and 512
processors.  Wall-clock time scales inversely with number of processors,
indicating near-optimal parallelism; timing analyses are presented in
\citet{barranco04a}.   

\section{RESULTS OF NUMERICAL SIMULATIONS}

In the following presentation of the numerical simulations, we define
a system of units such that $\Omega_0 = H_0 =  \rho_0 = 1$,
$c_{s0} = \Omega_0H_0 = 1$, and $RT_0 = c_{s0}^2 = 1$.
With these units, the Mach number of a computed flow is
$\epsilon=\max(\tilde{v})$ and the Rossby number is
$Ro=(1/2)\max(\tilde{\omega}_z)$.
All of the simulations presented in this section were computed with
$\tau_{cool}~\rightarrow~\infty$.  We will explore the effects of cooling
in future work.

\subsection{Vortex initialization}

The key to initializing a 3D vortex is to guess a vorticity configuration that is as close
to equilibrium as possible so that the initial accelerations are small and the flow can
slowly relax to its true equilibrium; otherwise, the vortex will be ripped apart by large
accelerations.  As discussed in Section 2.1, \citet{ms71} and \citet{kida81} analytically
determined exact steady-state solutions for 2D elliptical vortices of uniform vorticity
embedded in a linear background shear.   We use these 2D solutions as the starting
point for constructing initial conditions for 3D vortices.  We set $\tilde{v}_z$ and
$\partial\tilde{v}_z/\partial t$ to be initially zero, which implies that the horizontal velocity
fields at different heights are initially uncoupled and can be initialized independently.
At each height, we construct a 2D elliptical vortex of constant $\tilde{\omega}_z$ whose
strength and shape satisfy equation \eqref{E:moore_saffman} so that
$\partial\mathbf{\tilde{v}_{\bot}}/\partial t = 0$.  The ellipses at different heights do not
necessarily have to have the same size, shape, or strength; in the following  subsections,
we will describe two different ways of ``stacking'' 2D elliptical vortices to create a 3D vortex.
The anticyclonic core of the vortex is surrounded by a larger elliptical ``halo'' of weak
cyclonic vorticity so that the net circulation in each horizontal plane equals zero:
$\Gamma\equiv\int_A\tilde{\omega}_z dxdy = 0$.  We define a 2D streamfunction
$\tilde{\psi}$ such that:
$\mathbf{\tilde{v}_{\bot}}\equiv\mathbf{\nabla_{\bot}}\times\tilde{\psi}\mathbf{\hat{z}}$ and
$\tilde{\omega}_z = -\nabla^2_{\bot}\tilde{\psi}$.  Thus, once $\tilde{\omega}_z$ is
initialized at each height, we invert a 2D Poisson operator to obtain the streamfunction,
which in turn generates the horizontal velocity field.  The 2D pressure field can be found by
taking the horizontal divergence of the horizontal momentum equation and inverting another
2D Poisson operator.  The pressure will likely have a nonzero vertical gradient, which if left
unbalanced, will lead to large vertical accelerations.  We still have one more degree of
freedom: we initialize the temperature field so that the buoyancy force exactly balances the
vertical pressure force:
\begin{equation}
\tilde{T} = \frac{T_0}{\Omega_0^2z}\frac{\partial(\tilde{p}/\bar{\rho})}{\partial z}.
\end{equation}
Thus, all three components of the momentum equation are exactly balanced and the
accelerations are all initially zero: $\partial\mathbf{\tilde{v}}/\partial t = 0$.
Of course, only under very special circumstances would this procedure just so happen to
also exactly balance the temperature equation.  In general, $\partial\tilde{T}/\partial t\ne 0$
initially, leading to an immediate evolution of the temperature field.  Vertical motions
will then be generated by the temperature changes through the buoyancy force,
and these vertical velocities will then couple the horizontal motions at different heights.

\subsection{Tall, columnar vortex}

\begin{figure}
\epsscale{1.0}
\plotone{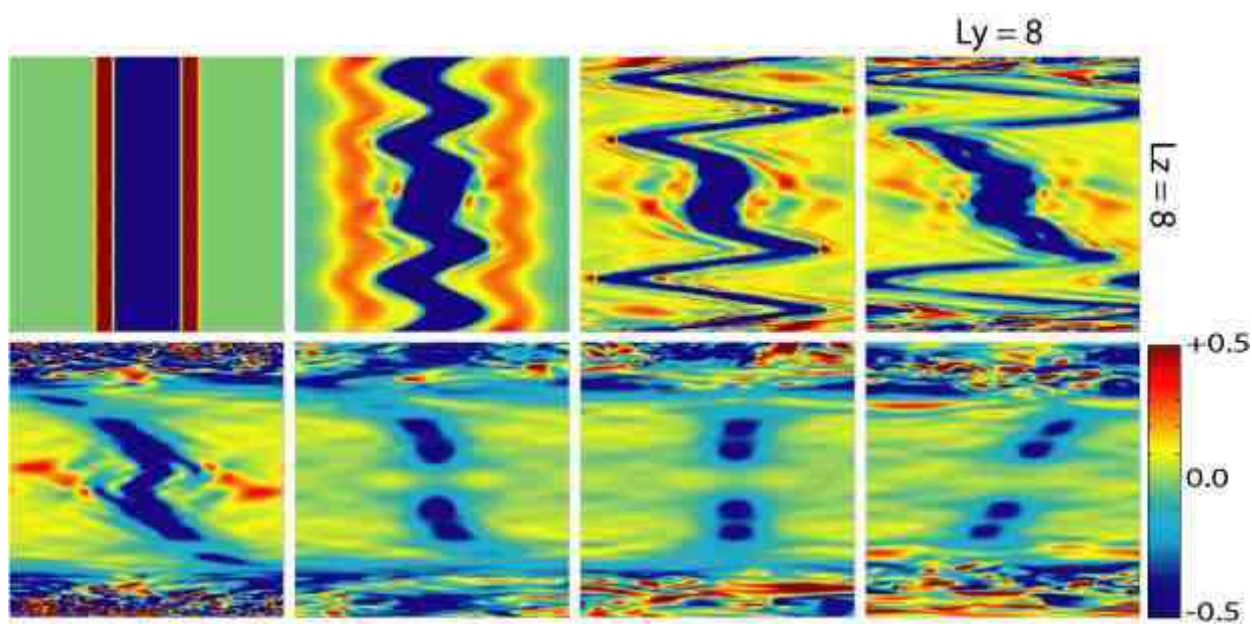}
\caption{\label{F:tall_column} Tall, columnar vortex:
$y\!-\!z$ slices at $x\!=\!0$ of the $z$-component of vorticity
$\tilde{\omega}_z$.  Blue = anticyclonic vorticity; red = cyclonic vorticity.
The initial elliptical column had aspect ratio $\chi=4$ and vorticity
$\tilde{\omega}_z = 0.625$, corresponding to a Rossby number
$Ro = 0.3125$.  The times corresponding to each frame are:
$t/\tau_{orb} = 0.0, 2.1, 4.2, 6.4, 51, 102, 153, 204$. }
\end{figure}

For our first initial condition, we constructed the simplest 3D analog of a 2D vortex:
an ``infinite'' column created by stacking identical 2D elliptical vortices at every height.
The vertical vorticity $\tilde{\omega}_z$, horizontal velocity $\mathbf{\tilde{v}_{\bot}}$ and
enthalpy $\tilde{p}/\bar{\rho}$ were all independent of height, and the temperature
perturbation was set identically to zero inside and outside the vortex.   This initial condition
was an exact equilibrium solution of the momentum and temperature equations:
$\partial\mathbf{\tilde{v}}/\partial t = \partial\tilde{T}/\partial t = 0$.
However, it turned out that this quasi-2D vortex is unstable to small perturbations in 3D.

Figure~\ref{F:tall_column} shows the evolution of a perturbed columnar vortex.
The domain dimensions for the simulation were $(L_x,L_y,L_z)=(2,8,8)$,  and the
numbers of grid points/spectral coefficients along each direction were
$(N_x,N_y,N_z) = (64,256,256)$.  The vertical velocity was forced to vanish on the top and
bottom boundaries of the domain: $\tilde{v}_z(z\!=\!\pm 4) = 0$.
The minor and major axes (along the $x$ and $y$ directions, respectively) of the elliptical
column were $(\Lambda_x,\Lambda_y) = (0.5,2.0)$, corresponding to an aspect ratio of
$\chi=4$.  The magnitude of the anticyclonic vorticity in the core of the
vortex was $\tilde{\omega}_z = 0.625$, corresponding to a Rossby number
$Ro=0.3125$.  The maximum value of the initial horizontal speed was
$\tilde{v}_{\bot,max} = 0.12$.   In order to investigate the stability of the tall column,
we did not stack the 2D ellipses exactly on top of one another, but instead
offset the locations of their centers in the radial direction
so that the centerline of the column had coordinates:
$\{x_c,y_c\} = \{A\sin(2\pi mz/L_z),0\}$, where $A=0.1\Lambda_r=0.05$ and $m=4$.

Because there is initially no vertical velocity or vertical enthalpy
gradient, the different layers of the vortex do not communicate with each
other, and tend to drift apart, carried by the ambient shear.  As the
vortex lines are stretched, a restoring force is generated which tries to
vertically realign the vortex.  The restoring force is partially successful in
realigning a segment of the vortex around the midplane, but the parts of the
vortex away from the midplane are sheared away, leaving a vertically truncated
vortex straddling the midplane.  Later, it appears that this truncated vortex
goes through another instability and splits into two independent vortices
above and below the midplane.  These off-midplane vortices survive for well
over a hundred orbits around the disk.

From this simulation, it is clear that vertical stratification plays a key role in
the vortex dynamics.
The vertical component of the protostellar gravity
is $g_z = \Omega_0^2z$ and the Brunt-V\"{a}is\"{a}l\"{a} frequency is
$\omega_{BV}\approx\Omega_0|z|/H_0$.  Far from the midplane $|z|>H_0$
the gas in the disk is strongly stratified ($Fr\equiv\tau_{BV}/\tau_{vor}<1$),
which inhibits the development of vertical motions which would couple the
layers together.  Thus, far from the midplane, the different layers of the
vortex are free to drift apart, carried by the ambient shear.
In contrast, there is virtually no stratification in the immediate vicinity
of the midplane ($Fr > 1$), and a vertical flow rapidly develops to couple the layers
together and prevent the ambient shear from ripping apart the vortex.

\subsection{Finite-height cylindrical vortex}

\begin{figure}
\epsscale{1.0}
\plotone{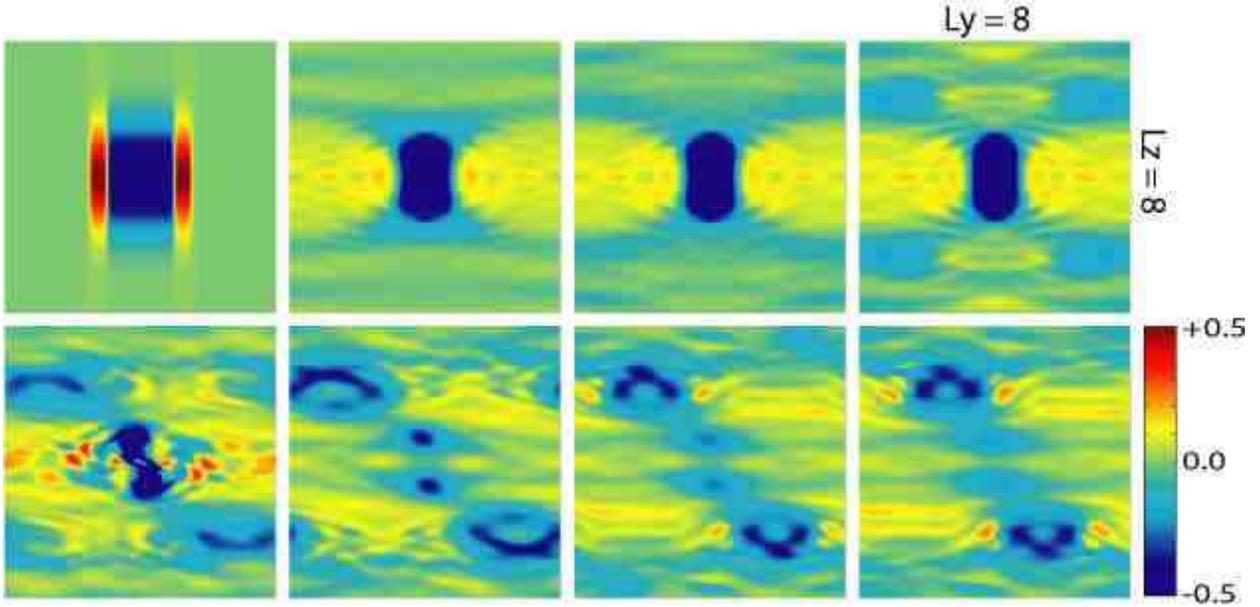}
\caption{\label{F:wz_yz} Finite-height cylindrical vortex:
$y\!-\!z$ slices at $x\!=\!0$ of the $z$-component of vorticity $\tilde{\omega}_z$.
Blue = anticyclonic vorticity, red = cyclonic vorticity.  
The initial elliptical cylinder had aspect ratio $\chi=4$ and vorticity
$\tilde{\omega}_z = 0.625$, corresponding to a Rossby number
$Ro = 0.3125$.   The time between frames is $\Delta t/\tau_{orb} \approx 60$.}
\end{figure}

\begin{figure}
\epsscale{1.0}
\plotone{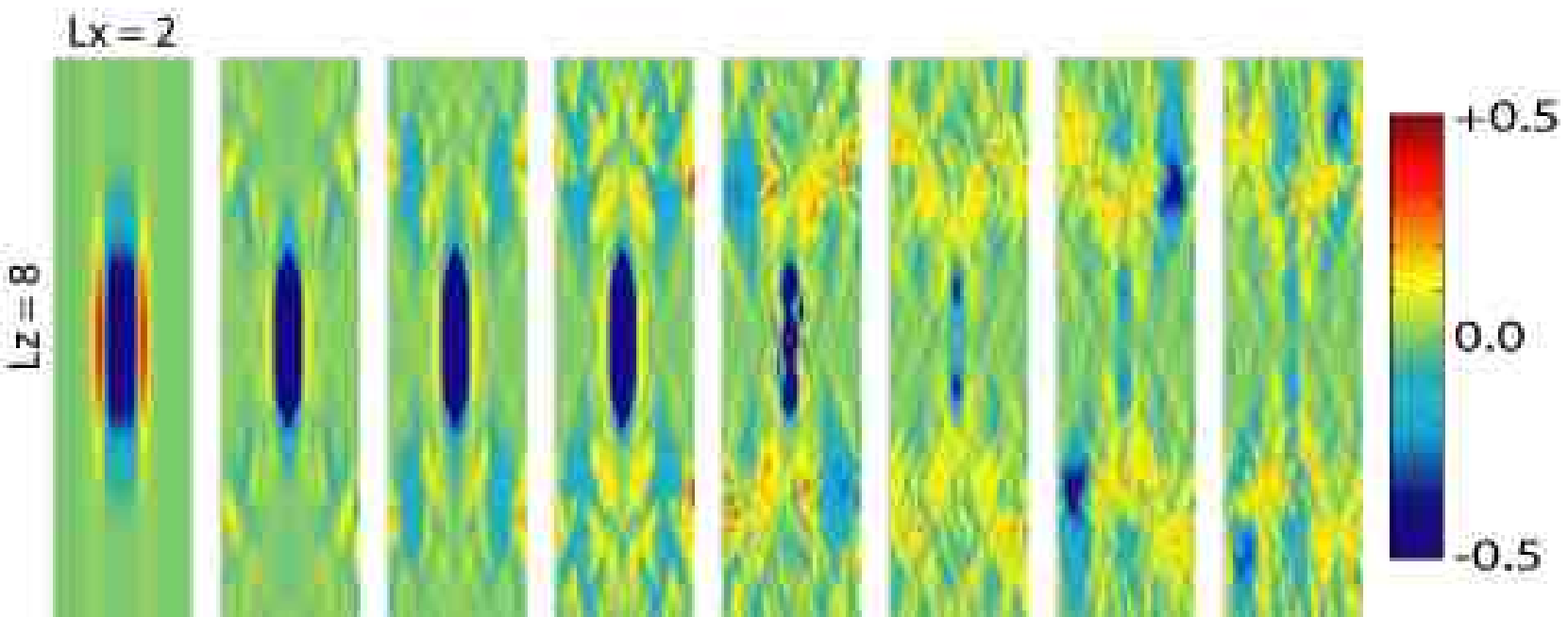}
\caption{\label{F:wz_xz} Finite-height cylindrical vortex:
$x\!-\!z$ slices at $y\!=\!0$ of the $z$-component of vorticity $\tilde{\omega}_z$.
Blue = anticyclonic vorticity, red = cyclonic vorticity.  
The initial elliptical cylinder had aspect ratio $\chi=4$ and vorticity
$\tilde{\omega}_z = 0.625$, corresponding to a Rossby number
$Ro = 0.3125$.   The time between frames is $\Delta t/\tau_{orb} \approx 60$.}
\end{figure}

\begin{figure}
\epsscale{1.0}
\plotone{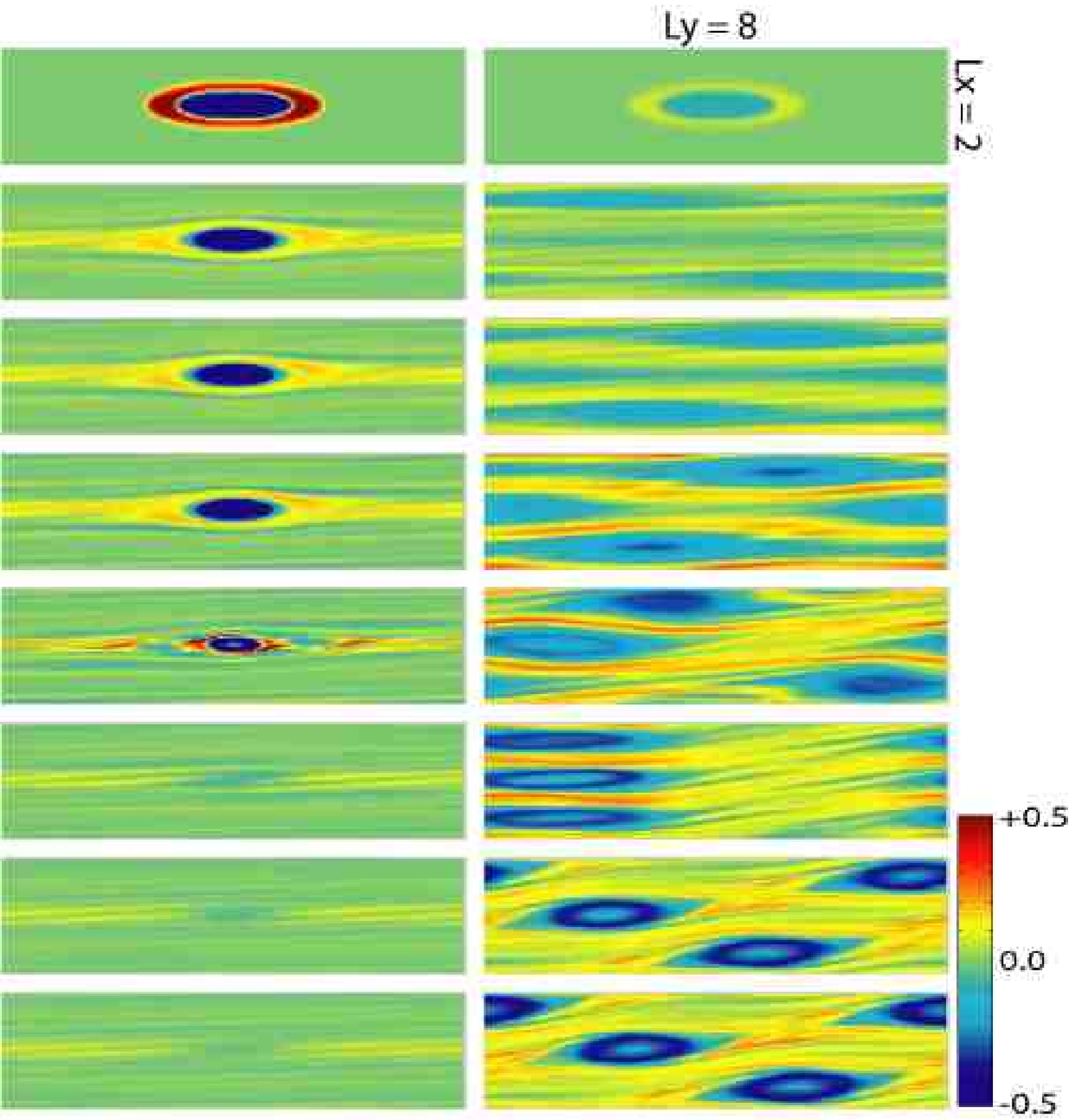}
\caption{\label{F:wz_xy} Finite-height cylindrical vortex:
$x\!-\!y$ slices at $z\!=\!0$ (first column) and $z\!=\!2$ (second column) of
the $z$-component of vorticity $\tilde{\omega}_z$.  
Blue = anticyclonic vorticity, red = cyclonic vorticity.  
The initial elliptical cylinder had aspect ratio $\chi=4$ and vorticity
$\tilde{\omega}_z = 0.625$, corresponding to a Rossby number
$Ro = 0.3125$.   The time between frames is $\Delta t/\tau_{orb} \approx 60$.}
\end{figure}

\begin{figure}
\epsscale{1.0}
\plotone{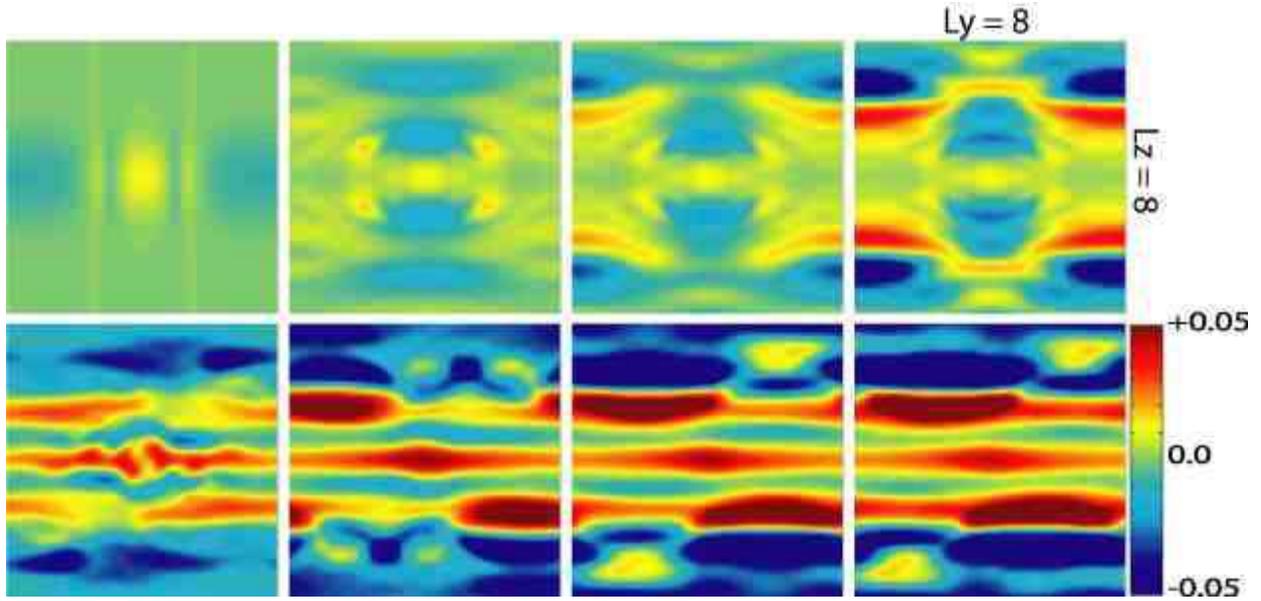}
\caption{\label{F:temp_yz} Finite-height cylindrical vortex:
$y\!-\!z$ slices at $x\!=\!0$ of the temperature perturbation $\tilde{T}$.
Blue = cool, red = warm.  
The initial elliptical cylinder had aspect ratio $\chi=4$ and vorticity
$\tilde{\omega}_z = 0.625$, corresponding to a Rossby number
$Ro = 0.3125$.   The time between frames is $\Delta t/\tau_{orb} \approx 60$.}
\end{figure}

\begin{figure}
\epsscale{1.0}
\plottwo{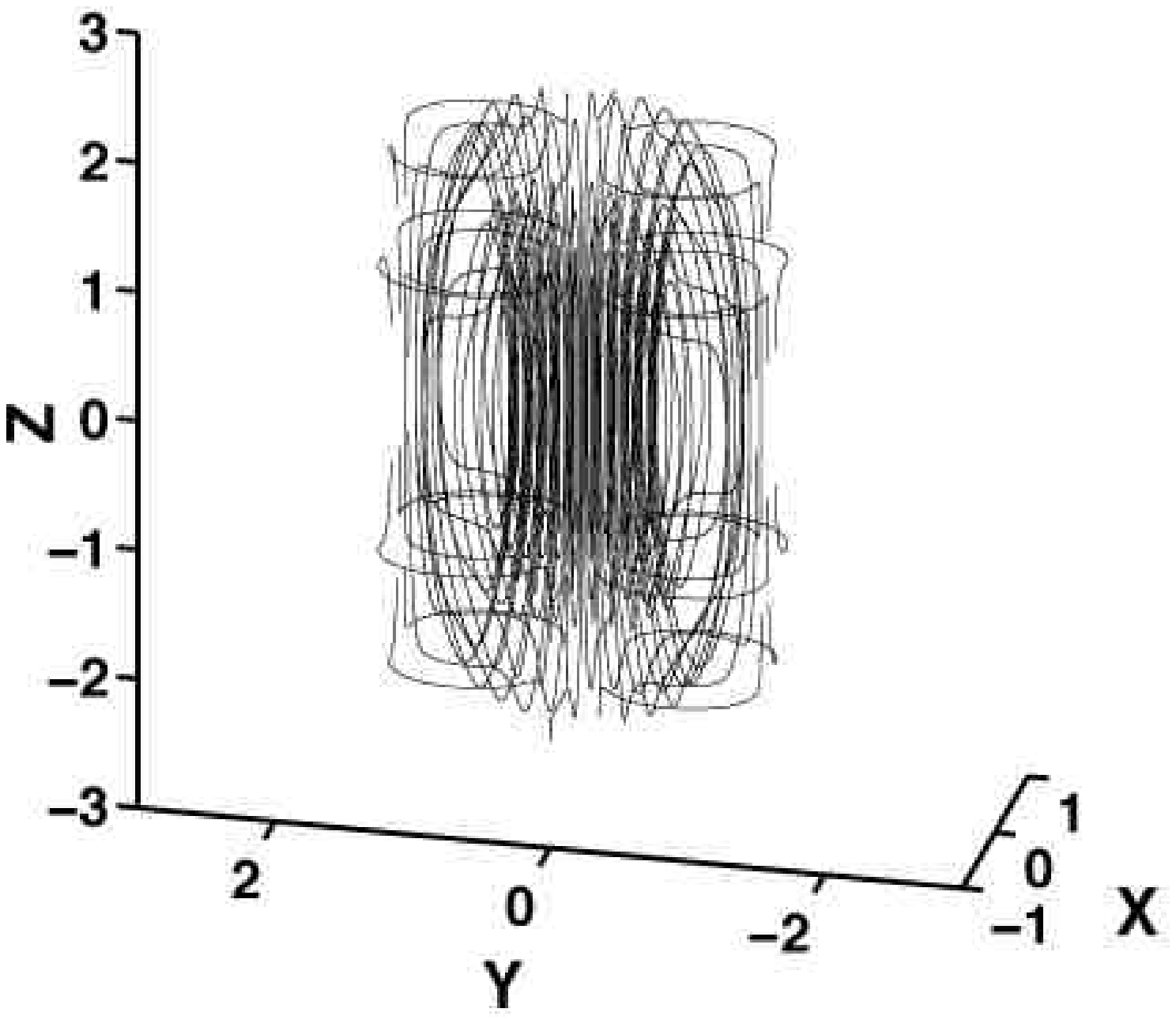}{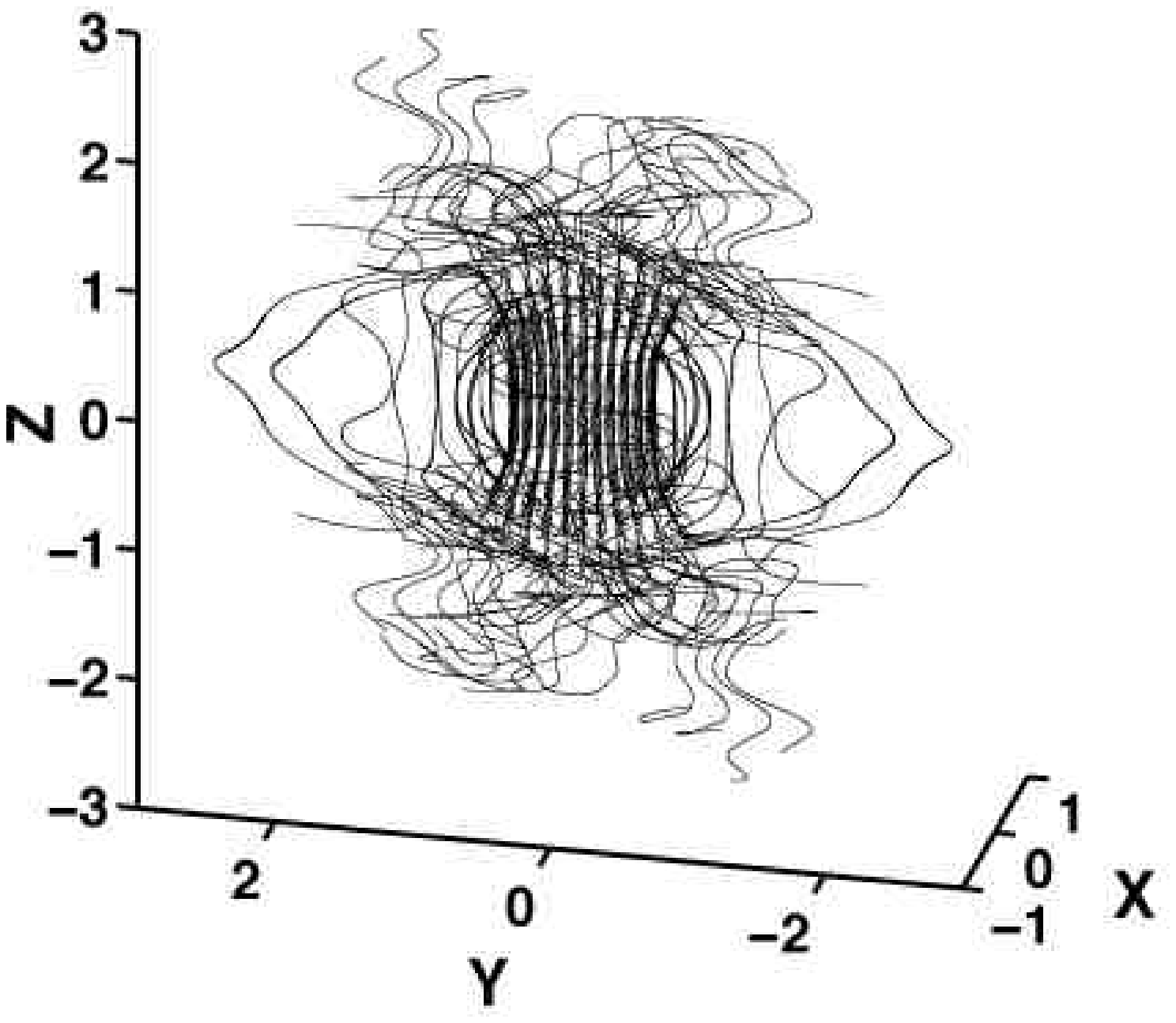}
\caption{\label{F:vortex_lines_sim} Vortex lines for finite-height cylindrical
vortex , at times $t/\tau_{orb}=0,170$.}
\end{figure}

\begin{figure}
\epsscale{0.60}
\plotone{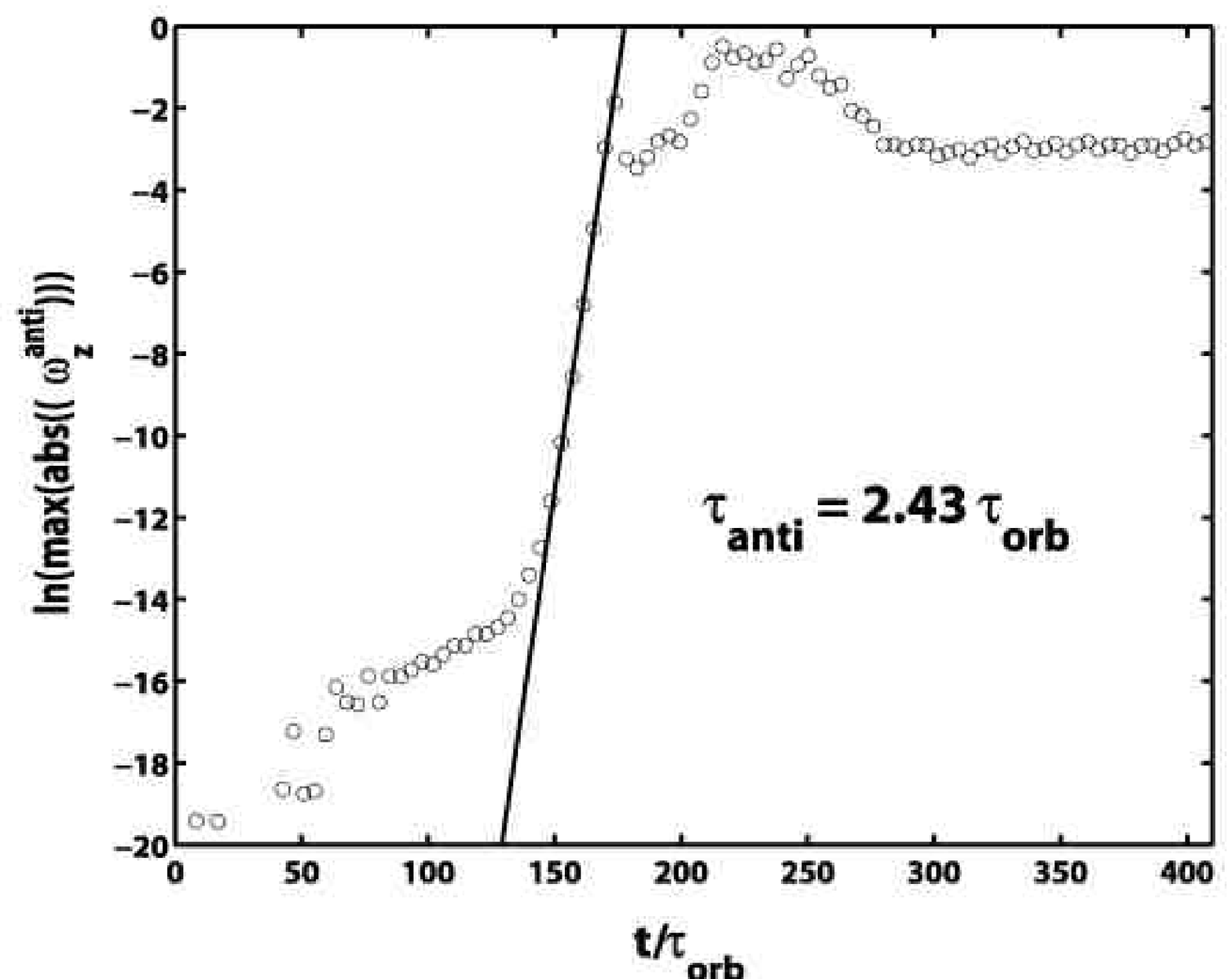}
\caption{\label{F:growth} Growth of unstable antisymmetric linear
eigenmode.  Because the initial condition was symmetric with respect to the
midplane, the instability could not be manifested until round-off error
populated the antisymmetric modes.  Eventually, the eigenmode emerges out of
the noise, and grows exponentially fast, with an $e$-folding time
of a few $\tau_{orb}$.}
\end{figure}

Motivated by the fact that tall, columnar vortices cannot maintain vertical coherence
over more than a couple scale heights,  we constructed a short, columnar vortex
that was vertically centered on the disk midplane.  We initialized a 2D ellipse
in the midplane whose minor and major axes were $(\Lambda_x,\Lambda_y) = (0.5,2.0)$,
corresponding to an aspect ratio of $\chi=4$.  The magnitude
of the vorticity in the core of the vortex was $\tilde{\omega}_z(z\!=\!0) = 0.625$ at the
midplane, corresponding to a Rossby number $Ro=0.3125$.  The maximum value of the
initial horizontal speed was $\tilde{v}_{\bot,max} = 0.12$.  The vorticity
was extended off the midplane according to a Gaussian profile:
\begin{equation}\label{E:Gaussian_vortex}
\tilde{\omega}_z(z) = \tilde{\omega}_z(0)e^{-z^2/2H_0^2},
\end{equation}
where the scale height of the vorticity was the same as the pressure scale
height.  For an anticyclone with $Ro\lesssim 1$, the inward Coriolis force is
somewhat more dominant than than the outward centrifugal force, and the
vortex must be a region of high pressure for horizontal equilibrium
(see Figure \ref{F:force_balance}).  The
high-pressure core extends only over a finite height, so that there is a
vertical pressure force away from the midplane.  In order for the vortex to be
in vertical equilibrium, there must be cool, dense lids which provide a
buoyancy force directed toward the midplane.

Figures~\ref{F:wz_yz} and \ref{F:wz_xz} show the time evolution of the
$z$-component of the vorticity in vertical slices $y\!-\!z$ at
$x\!=\!0$ and $x\!-\!z$ at $y\!=\!0$.  Figure \ref{F:wz_xy} shows the
$z$-component of the vorticity in two different horizontal slices $x\!-\!y$
at $z\!=\!0$ (first column) and $z\!=\!2$ (second column).
Figure~\ref{F:temp_yz} shows vertical slices $y\!-\!z$ at $x\!=\!0$
of the temperature perturbation.  The time between frames in all these figures
is $\Delta t/\tau_{orb}\approx 60$. These results were computed
with the infinite vertical domain version of the simulation: the horizontal
dimensions were $(L_x,L_y) = (2,8)$, and the vertical mapping parameter was
$L_z = 4$.  The numbers of spectral modes along each direction were
$(N_x,N_y,N_z) = (64,256,256)$.  The three components of the momentum equation
were exactly satisfied initially, but the energy equation was out of
equilibrium from the start.  The temperature field slowly
evolved, generating a small vertical velocity which then coupled the
horizontal layers together.  After approximately a few dozen $\tau_{orb}$, the
vertically truncated vortex settled into a new, quasi-equilibrium
(see frames 2-4 in Figures~\ref{F:wz_yz}--\ref{F:temp_yz}) that changed very
little over the course of a couple hundred orbits through the disk.
Figure~\ref{F:vortex_lines_sim} shows vortex lines for the initial condition
and for the quasi-equilibrium steady-state at $t/\tau_{orb}=170$.

We thought we had indeed found a stable steady-state, but surprisingly the
vortex suffered a dramatic instability which ultimately resulted in the
complete destruction of the vortex in the midplane (see frame 5 in
Figures~\ref{F:wz_yz}--\ref{F:temp_yz}).  The initial condition was symmetric
with respect to the midplane, and the equations of motion \eqref{E:euler}
should have preserved this symmetry.  However, the instability clearly had an
antisymmetric component.  We decomposed the flow into its symmetric and
antisymmetric parts.  Figure~\ref{F:growth} shows the maximum absolute value
of the antisymmetric part of the $z$-component of vorticity as a function of
time.  Initially, it was very close to zero, but grew from numerical round-off
errors.  Eventually, a linear eigenmode emerged out of this numerical
antisymmetric noise.  The structure of the mode preserved its spatial form
for over ten orders of magnitude of growth, proving that it is indeed a linear
instability.  The $e$-folding time of the exponential growth was a few $\tau_{orb}$. 

Thus, although the instability itself
had a fast timescale, the vortex appeared to survive for a long time in
its quasi-equilibrium state because a sufficient amount of numerical noise
had to be generated in the antisymmetric modes to seed the eigenmode.
In simulations in which we enforced symmetry about the midplane of the disk,
the truncated vortex survived for longer periods of time, but eventually
succumbed to slower symmetric instabilities.  

\subsection{Off-midplane anticyclonic vortices and cyclonic azimuthal bands}

\begin{figure}
\epsscale{0.75}
\plotone{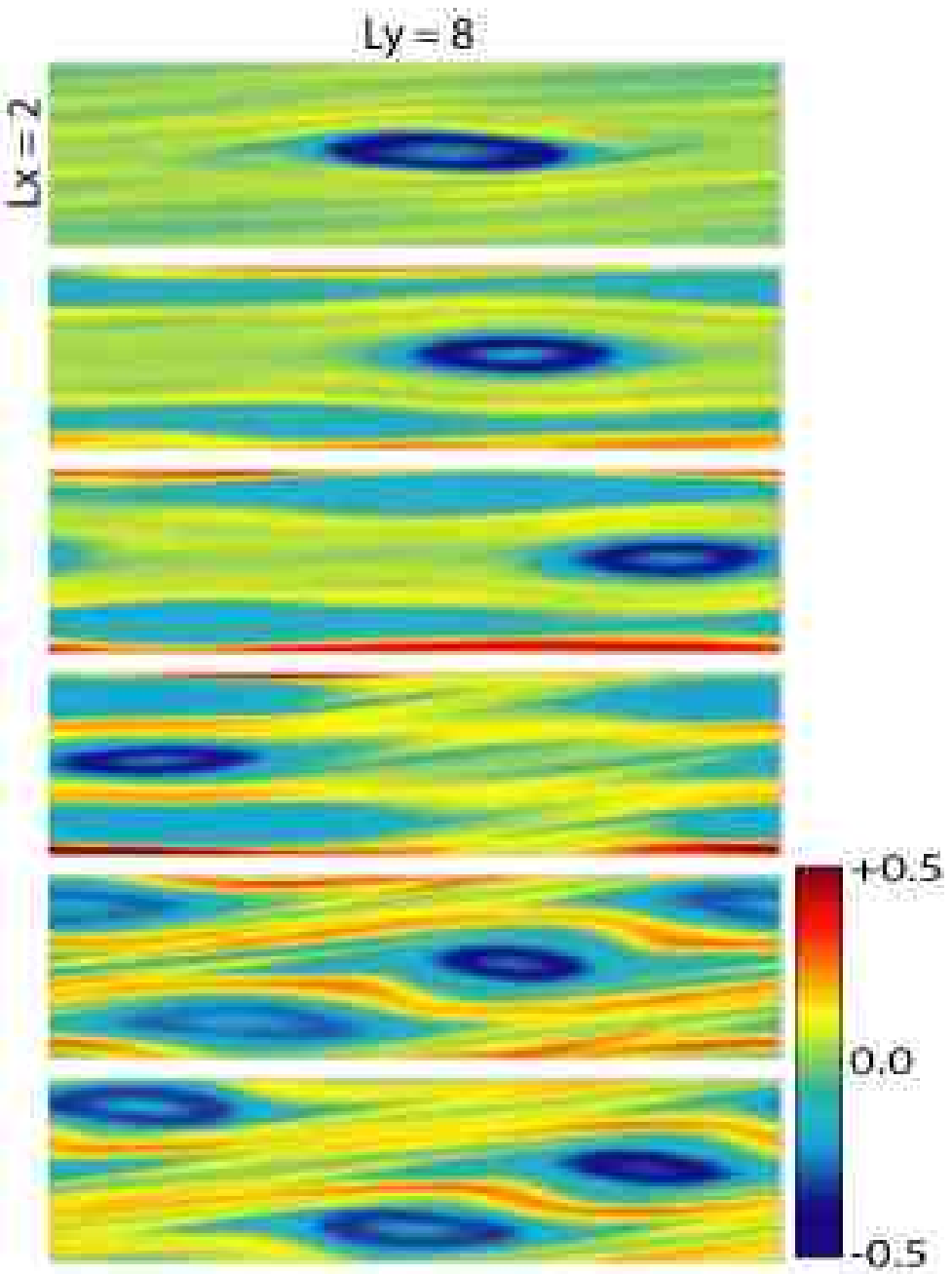}
\caption{\label{F:off_midplane} Simulation of an off-midplane vortex:
$x\!-\!y$ slice at $z=2$ of the $z$-component of vorticity $\tilde{\omega}_z$.
Blue = anticyclonic vorticity, red = cyclonic vorticity.  The time between frames is
$\Delta t/\tau_{orb} = 42$.  A single vortex from the
last frame of Figure~\ref{F:wz_xy} was isolated and used as the initial
condition for this run.  Surprisingly, the other vortices reformed.}
\end{figure}

A closer inspection of frames 4--8 in Figures~\ref{F:wz_yz}--\ref{F:temp_yz}
revealed coherent regions of anticyclonic vorticity a couple scale heights
above and below the midplane of the disk.  These new vortices formed
\textit{before} the vortex in the midplane succumbed to the antisymmetric
instability, and survived indefinitely in our simulations (over 400 orbits).
The highly stratified disk supports neutrally stable internal
gravity waves.  As a midplane vortex slowly relaxes to its quasi-equilibrium, it
oscillates and excites internal gravity waves which travel obliquely off the
midplane of the disk.
As the waves propagate off the midplane, conserving their flux of kinetic
energy, the velocity and temperature perturbations increase, the waves break,
and baroclinicity turns these perturbations into vorticity.
The anticyclonic shear of the disk stretches the cyclonic vorticity into
sheets or bands, whereas the anticyclonic vorticity perturbations roll-up
into new, coherent vortices.  Like the midplane-centered vortices, these
off-midplane vortices are also high-pressure anticyclones.
For a vortex in the upper half of the disk, the high-pressure
core is balanced by a cool, dense top lid which pushes downward, and a warm,
low-density bottom lid which pushes upward.  These lids can easily be seen
in frames 6--8 in Figure~\ref{F:temp_yz}.  The density and temperature of
these lids are maintained by the vertical motions within the vortex:
rising (falling) motion adiabatically cools (heats) the stably-stratified
fluid.

There are two intriguing features of these off-midplane vortices.
First, the three vortices approach one another and strongly interact,
stretching and squeezing each other.  Clearly, these are robust features
that are able to survive for hundreds of orbits despite strong perturbations.
One can see thin bands of cyclonic vorticity that keep the vortices separated;
the bands are most prominent at closest approach
(see the sixth frame down in column 2 of Figure~\ref{F:wz_xy}).
Second, the off-midplane vortices appear to be ``hollow'' in the sense that
the maximum vorticity does not occur at the center of the vortex, but on the
rim.  The Great Red Spot is probably the most famous example of a hollow
vortex; the center of the GRS may, in fact, be weakly counter-rotating
\citep{marcus93}.  It is still a mystery as to how the GRS maintains
its hollowness.  Numerical simulations of 2D hollow vortices show that they
are violently unstable; the low-vorticity fluid in the core switches places
with the high-vorticity fluid on the rim, resulting in a vortex whose
vorticity profile is centrally peaked \citep{youssef00, shetty04}.
Here in our 3D simulations, hollow vortices are apparently stable.

In order to get a better understanding on the structure of these off-midplane
vortices, we isolated one of them and used it as a new initial condition.
In this simulation, there is no vortex in the midplane, and except for the lone off-midplane
vortex at $z=2$, the rest of the domain is pure Keplerian with no initial pressure,
temperature, or density perturbations anywhere else.
Figure~\ref{F:off_midplane} shows the evolution of this isolated off-midplane vortex.
In frame 2 ($\approx 42\tau_{orb}$), one can clearly see anticyclonic bands forming. 
By frame 4 ($\approx 126\tau_{orb}$), the bands have rolled up and coalesced into
two new vortices.  The final state once again has 3 off-midplane vortices.
Most importantly, this simulation shows that that the vorticity for these new vortices
did not come from the redistribution of vorticity remnants from the destruction of a
midplane-centered vortex, but out of some new instability in the stratified region of the disk.
In a future paper, we will focus on the mechanisms for vorticity generation and the sources
of energy for these off-midplane vortices.

\subsection{Angular momentum transport}

\begin{figure}
\epsscale{1.0}
\plottwo{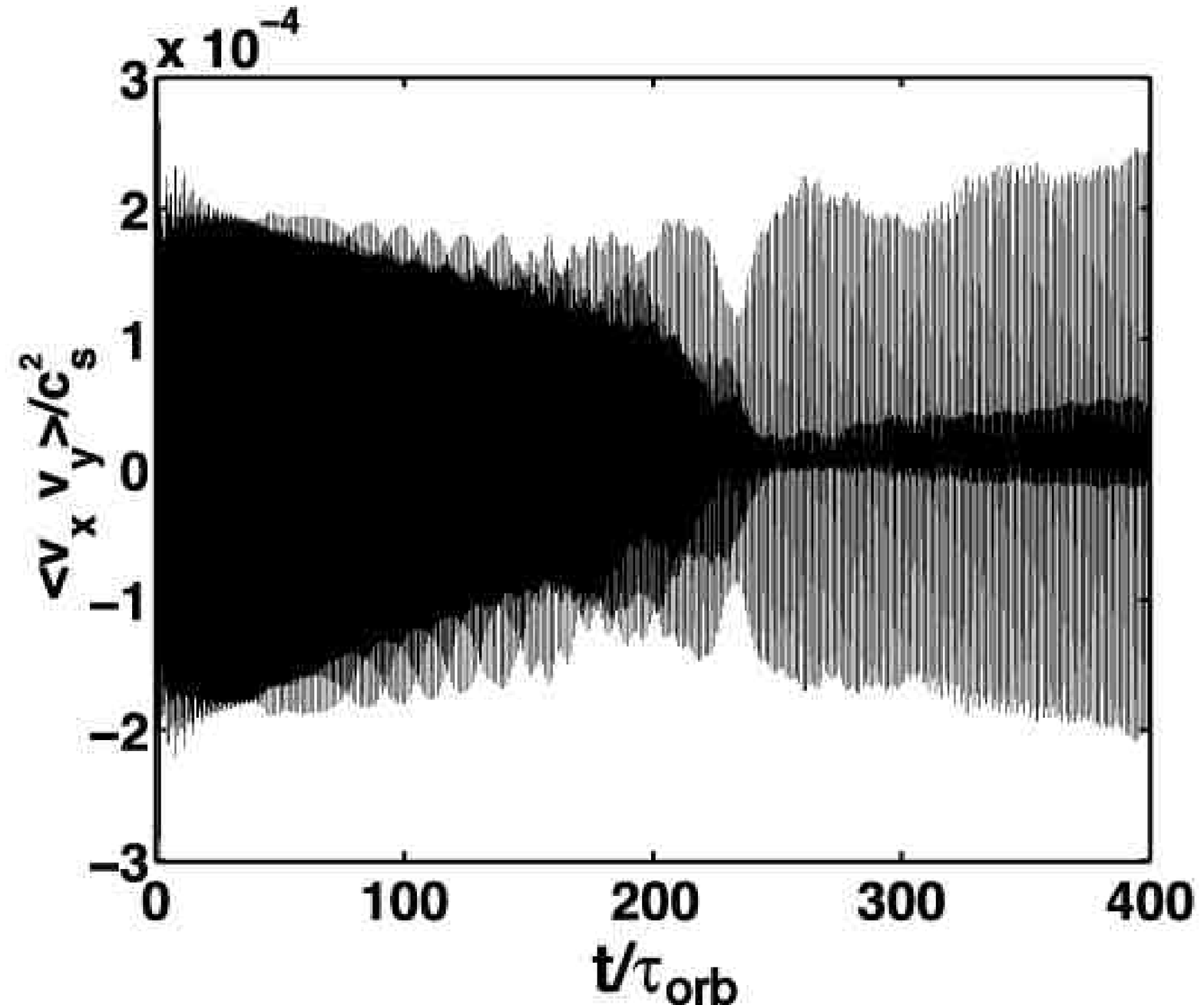}{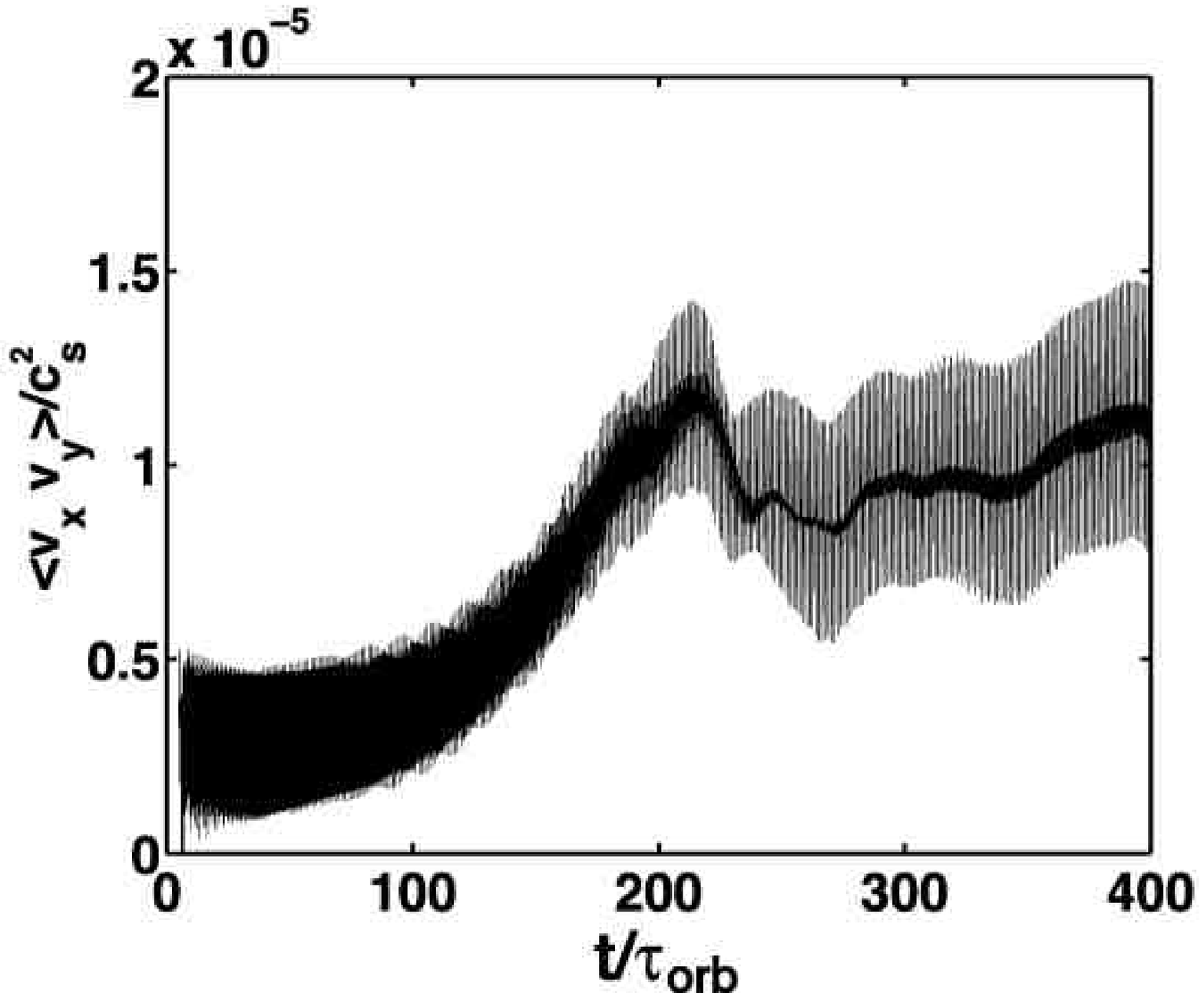}
\caption{\label{F:angmom} Velocity correlations as a function of time for the
full simulation of a finite-height midplane vortex settling into a
quasi-equilibrium, going unstable (at $t\approx 200\tau_{orb}$),
then new, off-midplane vortices forming.  The first panel shows the raw
velocity correlations, whereas the second panel shows the same data
smoothed with a window of $10\tau_{orb}$.}
\end{figure}

One of the prime motivations for interest in coherent vortices in
protoplanetary disks is whether they can transport sufficient angular momentum
outward so that mass can continue to accrete onto the growing protostar.
Following \citet{balbus98}, one can relate the correlations of the fluctuating
perturbation velocities (whether from turbulence or coherent vortices) to the
eddy-viscosity model: 
\begin{equation}
\alpha = \langle\tilde{v}_x\tilde{v}_y\rangle/c_s^2,
\end{equation}
where we have defined a density-weighted average $\langle\tilde{v}_x\tilde{v}_y\rangle\equiv\int\bar{\rho}\tilde{v}_x\tilde{v}_y dxdydz/\int\bar{\rho}dxdydz$.
For an isolated vortex that is fore-aft symmetric (\eg $\tilde{v}_x$ is antisymmetric across
the $x$-axis, symmetric across the $y$-axis and  $\tilde{v}_y$ is symmetric across the
$x$-axis, antisymmetric across the $y$-axis), one can
show that the velocity correlation defined above vanishes; these symmetries
must be broken in order to have non-zero transport.  However, if the disk is filled
with vortices, vortex-vortex interactions and mergers can break the fore-aft symmetry,
allowing for angular momentum transport.

In Figure~\ref{F:angmom}, we plot the velocity correlations for the long-term
simulation of a finite-height midplane vortex settling into a
quasi-equilibrium, going unstable, and new, off-midplane vortices forming.
In the first panel, we plot the instantaneous correlations, which are highly
time-dependent.  Note that the correlations are both positive (outward
transport of angular momentum) and negative (inward transport).
The strong positive and negative spikes in this raw data are due to the
interactions of the off-midplane vortices when they are at closest approach
(see frame 6 in column 2 of Figure~\ref{F:wz_xy}).
In the second panel, we present
a moving average (rectangular window with width $10\tau_{orb}$) of the
same data.  The smoothed correlations are always positive (outward transport of
angular momentum) and correspond to an average $\alpha\approx 10^{-5}$.

One must be careful when extrapolating the results of angular momentum transport in
a local patch of the disk to the global transport in a whole disk.
Our simulations are periodic in the azimuthal direction $y$, implying that
there is a periodic array of vortices around the protostar, with azimuthal
separation $L_y$.   However, vortices that are at the same radius tend to merge,
potentially reducing the effective rate of angular momentum transport.
The simulation shown in Figure~\ref{F:off_midplane} shows that new
vortices readily form in unoccupied radial bands.
Thus, there is a competition between mergers, which reduce the number of
vortices per radial band, and the formation mechanism that repopulates
empty bands.  In future simulations, we will extend the azimuthal domain
to investigate multiple vortices per radial band and explore this competition
between mergers and formation of new vortices. 

\section{SUMMARY \& FUTURE WORK}

This is the first \textit{ever} calculation of long-lived, robust 3D vortices in
protoplanetary disks.  The fundamental fluid dynamics are governed by three
key timescales (see Table~\ref{T:timescales}):
$\tau_{vor}$, the orbital period of gas around the vortex core;
$\tau_{orb}$, the orbital period of the vortex  around the protostar; and
$\tau_{BV}$, the Brunt-V\"{a}is\"{a}l\"{a} period, or period of vertical
oscillations (\eg internal gravity waves) in a stably-stratified atmosphere.
These three timescales are roughly of the same order because they are
all determined by the protostellar gravity.  However, the stratification in the disk
has a strong spatial dependence, making the dynamics more complicated.
The disk can roughly be divided into two regimes: a weakly
stratified region in the immediate vicinity of the midplane ($|z|<H_0$,
$Fr\equiv\tau_{BV}/\tau_{vor}>1$), and a strongly stratified region away from
the midplane ($|z|>H_0$, $Fr<1$).  Our numerical simulations have
confirmed that stratification is a crucial ingredient for long-lived
coherent vortices.

Earlier, 2D studies of vortices in disks
\citep[\eg][]{aw95,bracco98,godon99b,godon00} could not address the role
of stratification.  One way of interpreting their models is that a 2D vortex
is a cross-section (in the midplane) of a tall column that penetrates
through all the layers of the disk.  Our numerical simulations (see
Figure~\ref{F:tall_column}) show that such columns, although they are
\textit{exact} steady-state equilibrium solutions to the 3D momentum and
temperature equations, are unstable to 3D perturbations because stratification
inhibits vertical motions which are needed to couple the layers together and
lock the vertical alignment.  Another possible interpretation of 2D analyses is
that they capture the dynamics of vortices that are confined to the midplane of
the disk.  We initialized finite-height, cylindrical vortices straddling the
midplane of the disk and allowed them to evolve into quasi-equilibria.
These vortices were high-pressure anticyclones in which the inward Coriolis
force balanced the outward pressure force.  In the vertical direction,
the vortex needed cool, dense lids to exert a buoyant force toward the
midplane to balance the pressure force away from the midplane (see
Figure~\ref{F:force_balance}).  Long-term simulations of such vortices
revealed that they too are unstable equilibria.  We isolated an
antisymmetric (with respect to the midplane) eigenmode that grew with an
$e$-folding time of only a few orbital periods around the protostar. 
These vortices only seemed to last a long time because our initial conditions
were symmetric with respect to the midplane, and the antisymmetric eigenmode
was seeded only with numerical round-off error.

We confess that when we started this work, we too believed vortices in
protoplanetary disks would be either quasi-2D ``infinite'' columns,
or short truncated cylindrical vortices confined to the midplane of the disk.
Surprisingly, our simulations revealed an unexpected third possibility:
off-midplane vortices a few scale heights above and below the midplane.
These vortices are also high pressure anticyclones.  For a vortex in the upper half of the disk,
the high-pressure core is balanced by a cool, dense top lid which pushes downward, and a warm,
low-density bottom lid which pushes upward.  The density and temperature of
these lids are maintained by the vertical motions within the vortex:
rising (falling) motion adiabatically cools (heats) a stably-stratified fluid
(see Figure~\ref{F:force_balance}b).
Simulations with no vertical gravity and/or those with barotropic equations of state
will fail to find such vortices.  Also, numerical calculations must have sufficient
resolution to capture the thermal lids, which typically have thicknesses
much smaller than the local pressure scale height.

Unlike the tall, columnar vortices and the finite-height cylindrical midplane
vortices which we put in the disk by hand, the off-midplane vortices formed
naturally from perturbations in the disk.  For example, as a midplane vortex
oscillated and relaxed toward its quasi-equilibrium (\textit{not} the
cataclysmic motions as it succumbed to the antisymmetric instability), it
excited internal gravity waves which propagated away from the midplane,
steepened, and broke, creating vorticity (a baroclinic effect).  Regions of
anticyclonic vorticity rolled-up and coalesced into new anticyclones, whereas
regions of cyclonic vorticity were stretched into thin azimuthal bands and
sheets by the ambient shear.  In simulations where we had only one
off-midplane vortex as an initial condition, new vortices reformed,
filling every available radial band. 
These off-midplane vortices are
robust, surviving indefinitely in our simulations (over 400 orbits)
despite countless close encounters and interactions with other vortices.

The symmetries of an isolated vortex preclude significant angular momentum transport.
On the other hand, if the disk is filled with vortices that interact, merge and reform, those
symmetries are broken and angular momentum can be transported at moderate rates.
Thus, there is a competition between mergers of vortices that would reduce the
number of vortices in a given radial band, and the formation of new vortices in unoccupied
radial bands. 

In order to understand how vortices are created and maintained, and whether they
are isolated or fill the disk,  it is necessary to determine the sources of energy and
vorticity.  \citet{klahr03} demonstrated that a globally unstable radial entropy
gradient in a protoplanetary disk excited Rossby waves which also broke into
large-scale vortices.   Because the length scale of the entropy gradient
was of order $r$,  the horizontal scales of the
vortices were much larger than the thickness of the disk, and so the
fluid velocities were supersonic.  Acoustic waves and shocks rapidly drain the
kinetic energy from such vortices.  We plan to extend the radial domain
in our simulations and relax the shearing box boundary conditions so that
we may investigate radial gradients as well.

Coherent vortices located above and below the midplane will significantly
affect the way dust settles into the midplane.  Two-dimensional studies have
shown that vortices are very efficient at capturing and concentrating dust
particles \citep{barge95,tanga96,fuentemarcos01}.  If the vortices are located
off the midplane, will the dust grains be inhibited from settling into a thin,
dense sublayer?  If the vortices have a significant vertical velocity, 
will they be able to sweep up grains that have already settled into the midplane?
We propose that the trapping of dust in off-midplane vortices is analogous to the
formation of hail in the Earth's atmosphere; turbulent vertical velocities prevent
grains from falling out of the vortices until they have grown to some critical mass.
In future work, we plan to incorporate simple Lagrangian tracking of dust grains
\citep{fuentemarcos02}, as well as two-fluid models \citep{cuzzi93,johansen04}.

\acknowledgments

P.S.M. thanks the support of NASA Grant NAG5-10664 and NSF Grant AST-0098465.
J.A.B. thanks the NSF for support via a Graduate Student Fellowship while at Berkeley,
and now with an Astronomy \& Astrophysics Postdoctoral Fellowship (NSF Grant AST-0302146).
He also thanks the support of the Kavli Institute for Theoretical Physics through
NSF Grant PHY-9907949.
Computations were carried out at the San Diego Supercomputer Center using an NPACI award.
The authors would like to thank Berkeley graduate student Xylar Asay-Davis
for preparing the animations that appear in the electronic version of this paper;
and Professors Eugene Chiang, Geoff Marcy, and Andrew Szeri for useful
comments on the early manuscript.

\bibliographystyle{apj}

\end{document}